\documentclass[twocolumn]{aastex61}
\received{July 1, 2016}
\revised{September 27, 2016}
\accepted{\today}
\submitjournal{ApJ}

\shorttitle{Superoutbursts of Three Dwarf Novae Discovered by GWAC}
\shortauthors{Wang et al.}

\begin{document}

\title{Photometric and spectroscopic Studies of Superoutbursts of Three Dwarf Novae Independently Identified by 
The SVOM/GWAC System in 2018}

\correspondingauthor{J. Wang}
\email{wj@bao.ac.cn}
\correspondingauthor{H. L. Li}
\email{lhl@nao.cas.cn}

\author{J. Wang}
\affil{Guangxi Key Laboratory for Relativistic Astrophysics, School of Physical Science and Technology, Guangxi University,
Nanning 530004, People's Republic of China}
\affil{Key Laboratory of Space Astronomy and Technology, National Astronomical Observatories, Chinese Academy of Sciences, Beijing
100101, China}

\author{H. L. Li}
\affil{Key Laboratory of Space Astronomy and Technology, National Astronomical Observatories, Chinese Academy of Sciences, Beijing
100101, China}

\author{L. P. Xin}
\affil{Key Laboratory of Space Astronomy and Technology, National Astronomical Observatories, Chinese Academy of Sciences, Beijing
100101, China}

\author{X. H. Han}
\affil{Key Laboratory of Space Astronomy and Technology, National Astronomical Observatories, Chinese Academy of Sciences, Beijing
100101, China}

\author{X. M. Meng}
\affil{Key Laboratory of Space Astronomy and Technology, National Astronomical Observatories, Chinese Academy of Sciences, Beijing
100101, China}

\author{T. G. Brink}
\affil{Department of Astronomy, University of California, Berkeley, CA 94720-3411, USA}

\author{H. B. Cai}
\affil{Key Laboratory of Space Astronomy and Technology, National Astronomical Observatories, Chinese Academy of Sciences, Beijing
100101, China}

\author{Z. G. Dai}
\affil{School of Astronomy and Space Science, Nanjing University, Nanjing, 210000, China}

\author{A. V. Filippenko}
\affil{Department of Astronomy, University of California, Berkeley, CA 94720-3411, USA}

\author{C. -H. Hsia}
\affil{State Key Laboratory of Lunar and Planetary Sciences, Macau University of Science and Technology, Taipa, Macau, 999078, China}

\author{L. Huang}
\affil{Key Laboratory of Space Astronomy and Technology, National Astronomical Observatories, Chinese Academy of Sciences, Beijing
100101, China}

\author{L. Jia}
\affil{Key Laboratory of Space Astronomy and Technology, National Astronomical Observatories, Chinese Academy of Sciences, Beijing
100101, China}

\author{G. W. Li}
\affil{Key Laboratory of Space Astronomy and Technology, National Astronomical Observatories, Chinese Academy of Sciences, Beijing
100101, China}

\author{Y. B. Li}
\affil{Yunnan Observatories, Chinese Academy of Sciences, Kunming 650011, China}
\affil{Center for Astronomical Mega-Science, Chinese Academy of Sciences, 20A Datun Road, Chaoyang District, Beijing 100101, China}
\affil{School of Astronomy and Space Science, University of Chinese Academy of Sciences, Beijing, China}

\author{E. W. Liang}
\affil{Guangxi Key Laboratory for Relativistic Astrophysics, School of Physical Science and Technology, Guangxi University,
Nanning 530004, People's Republic of China}

\author{X. M. Lu}
\affil{Key Laboratory of Space Astronomy and Technology, National Astronomical Observatories, Chinese Academy of Sciences, Beijing
100101, China}

\author{J. Mao}
\affil{Yunnan Observatories, Chinese Academy of Sciences, Kunming 650011, China}
\affil{Center for Astronomical Mega-Science, Chinese Academy of Sciences, 20A Datun Road, Chaoyang District, Beijing 100101, China}
\affil{Key Laboratory for the Structure and Evolution of Celestial Objects, Chinese Academy of Sciences, Kunming 650011, China}

\author{P. Qiu}
\affil{Key Laboratory of Optical Astronomy, National Astronomical Observatories, Chinese Academy of Sciences, Beijing
100101, China}

\author{Y. L. Qiu}
\affil{Key Laboratory of Space Astronomy and Technology, National Astronomical Observatories, Chinese Academy of Sciences, Beijing
100101, China}

\author{J. J. Ren}
\affil{Key Laboratory of Optical Astronomy, National Astronomical Observatories, Chinese Academy of Sciences, Beijing
100101, China}

\author{D. Turpin}
\affil{Key Laboratory of Space Astronomy and Technology, National Astronomical Observatories, Chinese Academy of Sciences, Beijing
100101, China}

\author{H. J. Wang}
\affil{Key Laboratory of Optical Astronomy, National Astronomical Observatories, Chinese Academy of Sciences, Beijing
100101, China}

\author{X. G. Wang}
\affil{Guangxi Key Laboratory for Relativistic Astrophysics, School of Physical Science and Technology, Guangxi University,
Nanning 530004, People's Republic of China}

\author{X. Y. Wang}
\affil{School of Astronomy and Space Science, Nanjing University, Nanjing, 210000, China}

\author{C. Wu}
\affil{Key Laboratory of Space Astronomy and Technology, National Astronomical Observatories, Chinese Academy of Sciences, Beijing
100101, China}
\affiliation{School of Astronomy and Space Science, University of Chinese Academy of Sciences, Beijing, China}

\author{Y. Xu}
\affil{Key Laboratory of Space Astronomy and Technology, National Astronomical Observatories, Chinese Academy of Sciences, Beijing
100101, China}
\affil{School of Astronomy and Space Science, University of Chinese Academy of Sciences, Beijing, China}

\author{J. Z. Yan}
\affil{Purple Mountain Observatory, Chinese Academy of Sciences, Nanjing, 210034, China}

\author{J. B. Zhang}
\affil{Key Laboratory of Optical Astronomy, National Astronomical Observatories, Chinese Academy of Sciences, Beijing
100101, China}

\author{W. Zheng}
\affil{Department of Astronomy, University of California, Berkeley, CA 94720-3411, USA}

\author{J. Y. Wei}
\affiliation{Key Laboratory of Space Astronomy and Technology, National Astronomical Observatories, Chinese Academy of Sciences, Beijing
100101, China}
\affiliation{School of Astronomy and Space Science, University of Chinese Academy of Sciences, Beijing, China}



\begin{abstract}

We report our photometric and spectroscopic follow-up observations of the superoutbursts of three dwarf novae
(GWAC\,180415A, GWAC\,181017A and GWAC\,181211A) 
identified independently by the Ground Wide-angle Cameras system, one of the ground-based instruments of the 
China-France SVOM mission.
Based on a combination of our photometry and that taken from the AAVSO, our period analysis of the superhumps 
enables us to determine the mass ratios to be 0.0967-0.1163, 0.1879-0.1883 and 0.0981-0.1173 for GWAC\,180415A, GWAC\,181017A and GWAC\,181211A, respectively. 
GWAC\,180415A can be firmly identified as a WZ sge-type dwarf novae due to its long duration ($\sim2$ weeks) multiple rebrightenings with amplitudes of 3-4 magnitudes, the early superhump associated with a double-wave modulation and the low mass ratio. The inferred low mass ratio and location in the $\varepsilon-P_{\mathrm{orb}}$ diagram suggest that GWAC\,181211A is a 
WZ sge-type dwarf novae candidate.
The measured Balmer decrements suggest the Balmer line emission is produced from an optical thick region in GWAC\,180415A and GWAC\,181017A, and from an optical thin region in GWAC\,181211A. 

\end{abstract}




\section{Introduction} \label{sec:intro}

Cataclysmic variables (CVs) are semi-detached binaries that consist of a white dwarf (WD, the primary) 
and a low-mass star (the secondary or donor); see Waner 
(1995) for a comprehensive review. The secondary is 
either a low-mass main sequence star or a brown dwarf (BD). 
An accretion disk is formed around the WD from matter lost by the donor
through the inner Lagrangian point, 
resulting in an outburst due to a thermal instability in the  disk (e,g, Meyer \& Meyer-Hofmeister 1981;
Osaki 1989, 1996; Lasota 2001). Because of their bright outbursts, CVs are important optical transients for 
on-going and future ground-based time-domain surveys (e.g., Vestrand et al. 2004; Burd et al. 2015; Ivezic et al. 2008; Shappee et al. 2014; 
Chambers et al. 2016; Wei et al. 2016; Tonry et al. 2018; Graham et al. 2019). 
For instance, Breedt et al. (2014) released a catalog of 1000 CVs detected by the Catalina Real-Time Transient Survey (CRTS).

With U Geminorum (U Gem) star as a prototype,  Dwarf Novae (DNe) having relatively small mass-transfer rate are an 
important subset of CVs.  
Based on their outburst phenomena and parameters (e.g., orbital period, outburst amplitude and 
outburst frequency), there are mainly three sub-types of DNe. Among these, the SU UMa-type DNe  exhibit both frequent short outbursts 
lasting a few days and occasional superoutbursts with a duration of a couple of weeks. The orbital period
of the SU UMa-type DNe are typically short, usually below the well-known 2-3 hours period gap (e.g., Ritter \& Kolb 2003;
Wu et al. 1995). 
In addition, a superhump with a period longer than the orbital period by a few percent is identified in 
the SU UMa-type DNe without exception. The superhump is well explained by the dynamical precession of a quasi-elliptic disk
due to an expansion beyond the 3:1 resonance radius (e.g., Vogt 1982, Whitehurst 1988, Oskai 1989, Osaki \& Kato 2013).

The so-called WZ Sge-type objects are the SU UMa-type DNe with extremely long intervals between superoutbursts that is related to mass transfer (see a comprehensive review by Kato 2015). 
In contrast to other SU UMa-type DNe, the WZ Sge-type DNe show 
superoutbursts with amplitudes of 6-8 mag, small mass ratio  and extremely short period. Some WZ Sge-type DNe exhibit multiple rebrightenings after a dip (e.g., Imada et al. 2006;
Patterson et al. 2002; Meyer \& Meyer-Hofmeister 2015).    
The orbital period of a majority of WZ Sge-type DNe is $<$0.06 day ($\simeq87$ minutes),
and the mass ratio $q$ is typically smaller than 0.1, which implies a degenerate BD as the companion. In fact,
the evolutionary model of CVs predicts that $\sim40-70\%$ of CVs should have a BD 
companion and have passed the orbital period minimum, i.e., ``period bouncer'' (e.g., Kolb 1993; Howell et al. 1997;
Gansicke et al. 2009; Knigge et al. 2011; Goliasch \& Nelson 2015; Pala et al. 2018).
This predicted fraction is, however, much larger than the value ($\sim15\%$) determined from observations (e.g., Savoury et al. 2011), 
although a few of BD secondary stars have been indeed been identified by infrared spectroscopy in some previous studies 
(e.g., Ciardi et al. 1998; Littlefair et al. 2000, 2003, 2013; Howell \& Ciardi 2001; Mennickent et al. 2004; Aviles et al. 2010; Longstaff et al. 2019). 
A donor with a mass as low as $0.0436\pm0.0020M_\odot$ was spectroscopically identified in the eclipsing CV 
SDSS\,J105754.25+275947.5 by McAllister et al. (2017). 
The extreme low mass ratio makes WZ Sge-type DNe special in two ways: 1) the appearance of an inclination-dependent
early superhump caused by the 2:1 resonance (e.g., Lin \& Papaloizou 1979; Osaki \& Myer 2003); and 2) the  ``period bounce'' after passing the period minimum.

Here we report our follow-up photometry and spectroscopy of three DN's superoutbursts first detected by
the ASAS-SN survey (Shappee et al. 2014) but independently identified in 2018 by the 
Ground Wide-Angle Cameras system
(GWAC; Wei et al. 2016; Turpin et al. 2019), enabling us to identify 
two of them as new WZ Sge-type objects. 
The paper is organized as follows. Section 2 describes the designation and current status of the GWAC system.
The discovery and follow-up observations, along with data reduction, are presented in 
Sections 3. Sections 4 shows the observational results, paying attention to our period analysis.  
We discuss the implications in Section 5.

\section{GWAC System} \label{sec:style}

As one of the ground facilities of SVOM\footnote{SVOM is a China-France satellite mission dedicated to the detection and 
study of Gamma-ray 
bursts (GRBs). Please see White Paper given in Wei et al. (2016) for details.}, GWAC is 
designed not only to observe the prompt emission of GRBs in optical bands, but also to independently detect optical 
transients with a high cadence.
The total GWAC system will comprise 36 cameras, covering a total Field-of-View (FoV) of 5400$\mathrm{deg^2}$, and 
several additional follow-up telescopes. 
At the current stage,16 cameras have been deployed at Xinglong Observatory, National Astronomical
Observatories (NAOC). Thanks to the large sky coverage of the system, each GWAC camera is capable of independently searching for optical transients in the sky. Dedicated algorithms were developed to reduce data and to identify triggers promptly.  
The follow-up telescopes that are deployed beside the GWAC cameras include two 60cm telescopes (GWAC-F60A and GWAC-F60B) and one 30cm telescope (GWAC-F30, Xin et al. 2019). The 
follow-up deep imaging and spectroscopy can be carried out through Target of Opportunity observations by the NAOC 2.16m telescope (Fan et al. 2016) at Xinglong observatory and by the Lijiang 2.4m telescope (LJT) at Gaomeigu observatory.  We refer the readers to Appendix for a detailed description of the GWAC system.


\section{Observations and Data Reduction} \label{subsec:tables}

\subsection{Independent Discovery by GWAC}

The log of the three optical transients, i.e., GWAC\,180415A, GWAC\,181017A and GWAC\,181211A, independently discovered by the GWAC's
cameras is presented in Table 1. The UT discovery time is listed in Column (2).
Column (7) and (8) respectively give the $G$-band brightness and distance of the quiescent counterparts,
extracted from the \it Gaia\rm\ DR2 catalog (Gaia Collaboration et al. 2018).  
The typical localization error determined from the GWAC images is about 2\arcsec\ in all three cases. 
The shape of the image profile of the three triggers is very similar to the point-spread-function (PSF) of the nearby bright objects, suggesting with high probability that they are not hot pixels.
The three transients did not show any apparent motion among the several consecutive images.   
There are also no known minor planets or comets within a radius of 15\arcmin. 
Also, no known variable stars or CVs can be found in SIMBAD around the transient positions within 1\arcmin.

Aperture photometry is carried out for each frame by a dedicated on-line pipeline.  
The flux calibration was based on the nearby stars listed in the USNO B1.0 catalog. As an illustration,  
Table 2 shows an example of the photometric results of GWAC\,180415A.
Column (1) lists the modified Julian date (MJD) at 
the beginning of the exposure. The calibrated magnitudes in $R$-band and corresponding
uncertainties are tabulated in Columns (2) and (3), respectively.

\begin{table*}[h!]
\renewcommand{\thetable}{\arabic{table}}
\centering
\caption{Log of the three optical transients independently discovered by the GWAC cameras} 
\footnotesize
\label{tab:decimal}
\begin{tabular}{cccccccc}
\tablewidth{0pt}
\hline
\hline
ID & UT Time &  R.A. & Dec &  Other name & Quiescent counterpart & $G$-band & $d$\\ 
 &   &  &   & & & mag & pc \\
(1)  &   (2) & (3) & (4) & (5) & (6) & (7) & (8) \\
\hline
GWAC\,20180415A & 15:30:19 & 12:08:57.3 &  19:16:54 & ASASSN-18fk\dotfill  & SDSS\,J120857.3+191656.5\dotfill & $19.564\pm0.005$ & $340.4^{+112.2}_{-68.2}$\\
GWAC\,20181017A & 13:55:00 & 02:25:06.4 &  08:06:39 & ASASSN-18xt\dotfill  & AllWISE\,J022506.37+080638.4 & $20.190\pm0.017$ & \dotfill \\
GWAC\,20181211A & 12:34:53 & 01:48:23.3 & -00:48:07 & ASASSN-18abn & SDSS\,J014823.29-004807.4\dotfill & $20.536\pm0.014$ & $572.7^{+593.5}_{-269.2}$ \\
\hline
\hline
\end{tabular}
\end{table*}

\begin{table}[h!]
\renewcommand{\thetable}{\arabic{table}}
\centering
\caption{GWAC photometric results of GWAC\,180415A in $R-$band} 
\label{tab:decimal}
\begin{tabular}{lccl}
\tablewidth{0pt}
\hline
\hline
MJD & Brightness & Error\\ 
  day   & mag & mag \\
(1)  &   (2) & (3) \\
\hline
58223.660700  &  14.81   &  0.38 \\
58223.662088  &  14.82   &  0.43 \\
58223.664345  &  15.17   &  0.37 \\
58223.664693  &  14.87   &  0.41 \\
58223.666777  &  14.93   &  0.38 \\
58223.668512  &  15.00   &  0.41 \\
58223.669206  &  15.22   &  0.37 \\
58223.669380  &  15.13   &  0.37 \\
58223.671811  &  15.20   &  0.38 \\
58223.674000  &  15.26   &  0.38 \\
\hline
\hline
\end{tabular}
\end{table}

\subsection{Follow-up Photometry and Data Reduction}

We monitored the optical brightness of the three transients in standard Johnson-Bessell $B$, $V$, $R$ and $I$-bands 
by using the GWAC-F60A/B telescopes and the 0.76 m Katzman Automatic Imaging Telescope (KAIT) at Lick Observatory after the discovery.  Both GWAC-F60A and GWAC-F60B telescopes, operated jointly by
NAOC and Guangxi University, are identical. Each telescope is equipped with an Andor 2048$\times$2048 CCD as a
detector mounted at the Cassegrian focus. This setup finally results in a FoV of 19\arcmin.

The exposure times are 20-90 seconds
depending on the brightness of the objects, the used filter and the weather condition. The typical seeing was 
2-3\arcsec\ during our observations.

Raw images taken with the GWAC-F60A/B telescopes were reduced by following the standard routine in
the IRAF\footnote{IRAF is distributed by the National Optical Astronomical Observatories,
which are operated by the Association of Universities for Research in
Astronomy, Inc., under cooperative agreement with the National Science
Foundation.} package,  including bias and flat-field corrections. 
The dark-current correction was ignored since its
impact on the photometry was negligible once the CCD was cooled down to -60\degr C.

After standard aperture photometry, absolute photometric calibration was performed based upon several nearby comparison stars 
whose magnitudes in the Johnson-Cousins system were transformed from the SDSS Data Release 14 catalog
through the Lupton (2005) transformation\footnote{http://www.sdss.org/dr6/algorithms/sdssUBVRITransform.html
\#Lupton2005.}. 
An automatic image-reduction pipeline (Ganeshalingam et al. 2010; Stahl
et al. 2019), including bias and flat-field corrections and an astrometric solution, 
was used to reduce the raw images taken by KAIT. 
PSF photometry was then performed by using the DAOPhot from the IDL Astronomy Users Library.

An example of the follow-up photometry without correction for Galactic extinction is listed in Table 3
for GWAC\,180415A. Column (1) lists the MJD at 
the beginning of the exposure. The filter is given in Column (2). The calibrated magnitudes and the corresponding
1$\sigma$ uncertainties are tabulated in Columns (3) and (4), respectively. Column (5) lists the used telescope.

\begin{table}[h!]
\renewcommand{\thetable}{\arabic{table}}
\centering
\caption{An example of multi-wavelength photometric results of the follow-up observations of GWAC180415A.} 
\label{tab:decimal}
\begin{tabular}{ccccc}
\tablewidth{0pt}
\hline
\hline
MJD & Filter & Brightness & Error & Telescope\\ 
  day &  & mag & mag &  \\
(1)  &   (2) & (3)  & (4) & (5)\\
\hline
58225.245424  & B &  14.94  &  0.05 & GWAC-F60A\\
58225.246134  & B &  15.04  &  0.05 & GWAC-F60A\\
58225.247553  & B &  15.00  &  0.05 & GWAC-F60A\\
58225.248263  & B &  15.01  &  0.05 & GWAC-F60A\\
58225.248972  & B &  14.99  &  0.05 & GWAC-F60A\\
58225.249681  & B &  15.11  &  0.04 & GWAC-F60A\\
58225.250390  & B &  15.02  &  0.04 & GWAC-F60A\\
58225.251100  & B &  15.07  &  0.04 & GWAC-F60A\\
58225.251809  & B &  15.09  &  0.04 & GWAC-F60A\\
58225.276204  & B &  15.12  &  0.06 & GWAC-F60A\\
\hline
\hline
\end{tabular}
\end{table}

\subsection{Follow-up Spectroscopic Observations and Data Reduction}

After the detection of each of the three transients, long-slit
spectroscopy was carried out in multiple epochs by the following telescopes: 
1) LJT at Yunnan Observatory, 2) NAOC 2.16m telescope at Xinglong Observatory, and 3) Shane 3m telescope at Lick Observatory.
A log of the spectroscopic observations is presented in Table 4.

\begin{table}[h!]
\renewcommand{\thetable}{\arabic{table}}
\centering
\caption{Log of spectroscopic observations}
\label{tab:decimal}
\tiny
\begin{tabular}{ccccc}
\tablewidth{0pt}
\hline
\hline
Transients & Date & Exposure & $\Delta\lambda$ & Telescope/Spectrograph \\
           &      &  seconds &    \AA          &   \\
 (1)  & (2) & (3) & (4) & (5) \\
\hline
GWAC\,180415A & 2018-04-16UT16:00 & 750  & 10 & LJT/YFOSC \\
              & 2018-04-16UT18:00 & 900  & 10 & LJT/YFOSC \\
              & 2018-04-20        & 2400 & 10 & 2.16m telescope/OMR \\
              & 2018-04-23        & 2400 & 10 & 2.16m telescope/OMR \\
              & 2018-05-03        & 1800 & 2.5 & 2.16m telescope/BFOSC \\
GWAC\,181017A & 2018-10-17        & 1800 & 10  & 2.16m telescope/OMR \\
GWAC\,181211A & 2018-12-12        & 1800 & 10  & 2.16m telescope/BFOSC \\
              & 2018-12-14        & 1200 & 10  & 2.16m telescope/BFOSC \\
              & 2018-12-19        & 1800 & 10  & 2.16m telescope/BFOSC \\
              & 2019-01-04        & 1800 & 12  & Shane telescope/Kast \\                   
\hline
\hline
\end{tabular}
\end{table}

The long-slit spectra taken by LJT use the Yunnan Faint Object Spectrograph and Camera (YFOSC; Fan et al. 2015) that 
is equipped with a back-illuminated 2048$\times$4096 E2V42-90 CCD as a detector. 
The grating G14 and a 2.5\arcsec\ slit oriented in the south-north direction were used. 
This setup results in a spectral resolution of $\sim$11\AA\ as measured from the sky 
emission lines and comparison lamps, and provides a wavelength coverage from 3250\AA\ to 7500\AA.
The wavelength calibration was carried out with the iron-argon comparison lamps.

The long-slit spectra taken with the NAOC 2.16m telescope use either the Optomechanics 
Research (OMR) Inc. Spectrograph or the
Beijing Faint Object Spectrograph and Camera (BFOSC).  The OMR spectrograph is equipped with a back-illuminated 
SPEC 1340$\times$400 pixels CCD as a detector.
The grating of 300$\mathrm{grooves\ mm^{-1}}$ blazed at 6000\AA\ and a slit of a width of 2\arcsec\ oriented 
in the south-north 
direction were used, giving a spectral resolution of
$\sim$10\AA. Helium-argon comparison lamps were adopted for the wavelength calibrations.
The BFOSC spectrograph is equipped with a back-illuminated E2V55-30 AIMO CCD. 
With a slit width of 1.8\arcsec\ oriented in the south-north direction, the spectral resolutions are 
$\sim10$\AA\ and $\sim2.5$\AA\ with gratings G4 and G8, respectively. 
The corresponding wavelength coverage are 3850-8000\AA\ and 5800-8000\AA.\ 
The wavelength calibrations were carried out with the iron-argon comparison lamps.

The spectrum of GWAC\,181211A on 4 January  2019 was taken with the Kast spectrograph mounted on the 3m Shane telescope at 
Lick Observatory. Grism 600/4310 was used on the blue side (1860 s) and grating
300/7500 was used on the red side (1800 s). 
The slit width was 2\arcsec, resulting in a 
spectral resolution of  $\sim5$\AA\ and $\sim12$\AA\ on the blue and red sides, respectively.

All the spectra were obtained as close to meridian as possible, to minimize the effects of atmospheric dispersion.
The Lick/Kast slit was oriented along the parallactic angle 209$\degr$ at an airmass of 1.4. 
In all but one case, the flux calibrations were carried out with observations of Kitt Peak National Observatory (KPNO)
standard stars (Massey et al. 1998); no standard stars were observed on 03 May 2018 when the GWAC180415A
spectrum was taken.

Standard procedures were adopted to reduce the two-dimensional spectra 
using the IRAF package, including bias subtraction and flat-field correction. 
All of the extracted one-dimensional spectra were then calibrated in wavelength and in flux by the corresponding
comparison lamp and standards, except for the GWAC\,180415A spectrum taken on 03 May 2018. The accuracy of wavelength calibration
is better than 1\AA\ for the OMR, YFOSC and Kast spectra, and better than 2\AA\ for the BFOSC spectra. 
The  telluric A-band (7600-7630 \AA) and B-band (around 6860\AA) features  
due to $\mathrm{O_2}$ molecules was removed from many (but not all) of the 
spectra using observations of the corresponding standard.

\section{Results and Analysis}

\subsection{Photometric light curves}

The combined multiwavelength light curves are respectively shown in Figures 1, 2 and 3 for 
GWAC\,180415A, GWAC\,181017A and GWAC\,181211A. 
In addition to the photometric results obtained from the GWAC cameras, GWAC-F60A/B telescopes and KAIT,
the results extracted from the American Association of Variable Star Observers (AAVSO) websit\footnote{https://www.aavso.org/.} 
are also overplotted. When compared to the brightness of the corresponding counterparts in the
quiescent state (i.e., Column (7) in Table 1), all three transients reveal a superoutburst with not only an amplitude $\geq6$ mag but also a 
duration longer than two weeks, typical of WZ Sge-type DNe. 
The observed superoutburst phenomenon is described for each transient as follows.

\subsubsection{GWAC\,180415A}

The very bright transient ASAS-SN\,18fk, with a peak of $V\approx$ 12 mag, was reported on 17 March 2018 by the 
ASAS-SN survey (Shappee et al. 2014). Unfortunately, that sky position was not covered by 
the GWAC cameras until 15 April 2018. 
The light curve (Fig. 1) shows a long-duration ($\sim2$ weeks) rebrightening consisting of
several outbursts with an amplitude of 3-4 mag, strongly suggesting that this object is a WZ Sge-type DN.
Our observations and resulted light curves are consistent with the study done by Pavlenko et al. (2019). 
ASAS-SN\,18fk has been observed immediately after the discovery by Pavlenko et al. (2019). 
Their observations were carried out with 18 telescopes located at 15 observatories 
during 70 nights in unfiltered light.  As same as our light curves, their light curve clearly shows six rebrightenings.

\subsubsection{GWAC\,181017A}

This transient with a brightness of 15.59 mag (ASAS-SN\,18xt) was first discovered by the ASAS-SN survey on 10 October 2018, earlier than the first detection by the GWAC system by 7 days. 
The exponential decay is well sampled for this transient by the GWAC-F60A telescope. We do not plot the photometric 
results given by the GWAC's cameras because the signal-to-noise ratio (S/N) is marginal for this transient.

\subsubsection{GWAC\,181211A}

A ``text-book'' light curve, including early raise, power-law decay, and subsequent exponential decay, 
has been obtained for this transient. The good sampling after the dip by the GWAC-F60A/B telescopes enable us 
to firmly exclude an existence of a rebrightening. Thanks to the early monitor by GWAC and GWAC-F60A/B telescopes,
the peak amplitude of the superoutburst is inferred to be $\sim6$ mag. 



\begin{figure}
\plotone{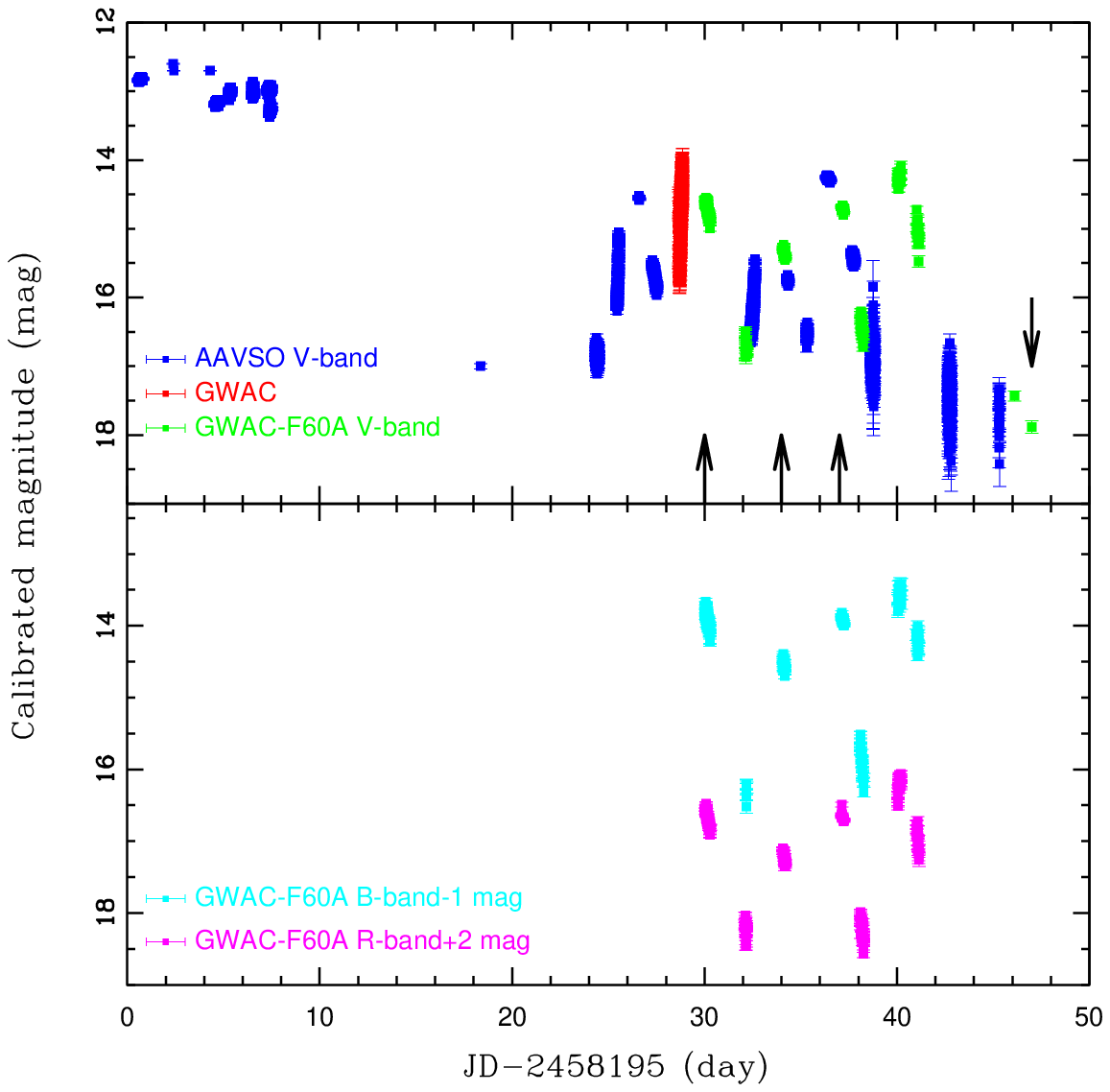}
\caption{The multi-wavelength light curves of GWAC\,180415A since the first detection reported by ASAS-SN survey. All the magnitudes are not corrected for the 
Galactic extinction. The overplotted vertical arrows mark the epochs when the spectra were obtained by us.  
}
\end{figure}

\begin{figure}
\plotone{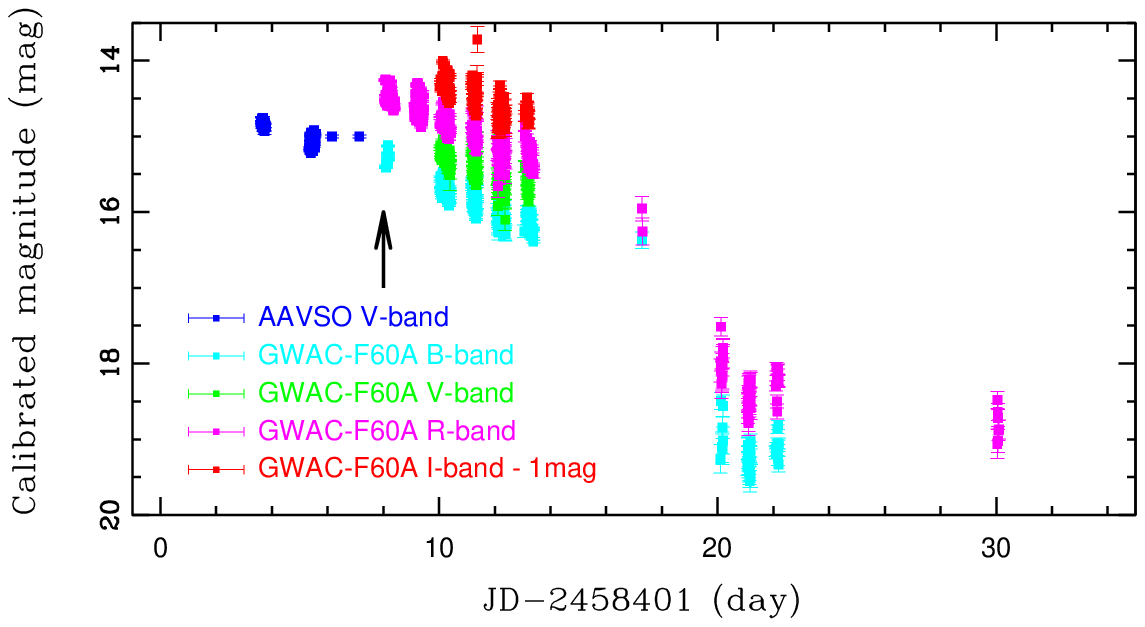}
\caption{The same as the Figure 1, but for GWAC\,181017A.}
\end{figure}

\begin{figure}
\plotone{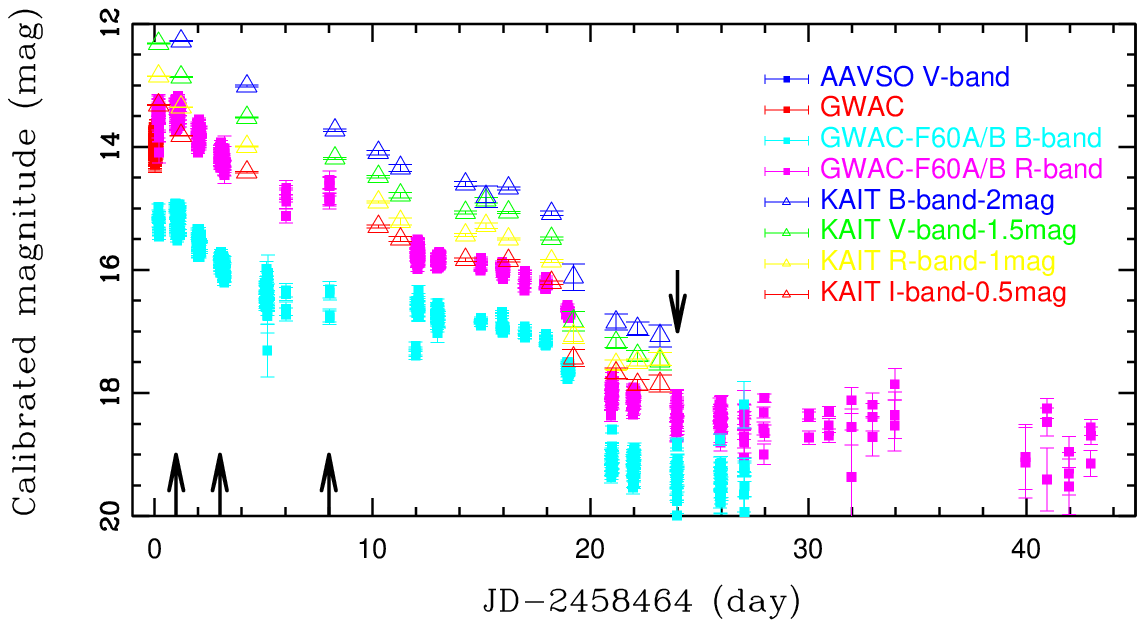}
\caption{The same as the Figure 1, but for GWAC\,181211A. }
\end{figure}

\subsection{Spectral evolution}

The spectral evolution of the three superoutbursts is shown in Figures 4-6. The 
epochs when these spectra were obtained are marked in Figures 1-3 by the vertical arrows.  
The blue continuum plus emission and absorption features are typical of a superoutburst of a DN 
(e.g., Wu et al. 2001; Baba et al. 2002; Zhao et al. 2006; Sheets et al. 2007; Hiroi et al. 2009; Breedt et al. 2014; Neustroev et al. 2017). 
The details are described for each DN as follows. 

\subsubsection{GWAC\,180415A}

The spectra of GWAC\,180415A were obtained in its rebrightening phase. 
A weak and possible double-peaked H$\alpha$ emission line can be identified
in all five spectra. The four flux-calibrated spectra are dominated by the high-order broad Balmer 
absorption lines at the blue end.
A transition from emission to absorption is revealed from the line profile of H$\beta$, which
clearly shows a broad absorption superposed with a narrow emission profile at the line center. The broad 
Balmer absorptions and superposed narrow emission lines are believed to be emitted from 
an optically thick disk and cool gas at the outer disk, respectively (e.g., Clark \& Bowyer 1984; Chen \& Lin 1989). 
The H$\alpha$ and H$\beta$ emission line profiles corrected by an absorption line described by a 
Gaussian function are illustrated in the bottom and middle rows in Figure 7 for 
two epochs (i.e., 16 and 23 April 2018).  Based on the correction of the absorption lines, 
the Balmer decrements H$\alpha$/H$\beta$ are measured to be $\sim1.4$ and $\sim1.5$; 
however, the uncertainties are relatively large because of the low S/N. The flat decrements suggest
that the Balmer emission lines are produced in an optically thick accretion disk in the rebrightening stage in GWAC\,180415A.

In addition, there are two notches at $\sim$4600\AA\ and $\sim$5200\AA\ in the first two spectra taken by LJT/YFOSC on 16 Apr 2018. 
Although without a convincing identification, we tentatively identify them as absorptions caused by 
$\mathrm{Fe^+}$ ions. 

There is no detectable 
\ion{He}{2}$\lambda4686$, \ion{He}{1}$\lambda$5016, \ion{He}{1}$\lambda$6678 or Bowen blend in the spectra. 
This is quite interesting because Pavlenko et al. (2019) argued that this object is potentially an intermediate 
polar (IP). The argument of an IP is a strong  22-min brightness modulation that is superimposed on superhumps in only 
the rebrightening and decline phases£¬which is however inconsistent with the observed spectra.  
It is known that IPs are typical of strong \ion{He}{2}$\lambda4686$ and Bowen lines (e.g., Negueruela et al. 2000; Saito et al. 2010). We admit that our spectra were all obtained in the outburst phase. A spectrum taken in quiescent state is helpful for 
further distinguishing. 
\rm

\subsubsection{GWAC\,181017A}

Only one spectrum was obtained for GWAC\,181017A on 17 October 2018. In addition to the 
H$\alpha$ emission line, the \ion{He}{2}$\lambda4686$ and \ion{C}{3}/\ion{N}{3}$\lambda\lambda4634-4651$ 
emission are clearly detected.   
We measure the H$\alpha$/H$\beta$ line ratio by direct integration, resulting in a rather flat value of $\sim2.0$. 
Again, the Balmer emission lines probably come from an optically thick accretion disk. The 
\ion{He}{2} emission line is believed to be produced by a hot chromsphere or coronas above the disk (e.g., Williams 1995; 
Sarty \& Wu 2006).  
  
\subsubsection{GWAC\,181211A}

The spectra of GWAC\,181211A are typical of a DN. The H$\alpha$ emission line resulting from the 
accretion disk exhibits a double-peaked profile in three out of the four cases. The \ion{He}{2}$\lambda4686$ and 
\ion{C}{3}/\ion{N}{3}$\lambda\lambda4634-4651$ line can be identified in the first two spectra (see the references cited above). 
The \ion{He}{1}$\lambda$5016+\ion{C}{4}/\ion{C}{3} emission can be marginally identified in the first spectrum taken one day after 
the onset of the superoutburst. The absorption-corrected H$\alpha$ and H$\beta$ emission line profiles are
shown in the top row of Figure 7 for the spectrum taken on 12 December 2018. The Balmer decrement is measured to be
large,  $\sim5.6$, suggesting optically thin condition for the Balmer line-emission region. However, once again, the uncertainty is large because of the low S/N and model-dependent measurement of H$\beta$.

\begin{figure}[ht!]
\plotone{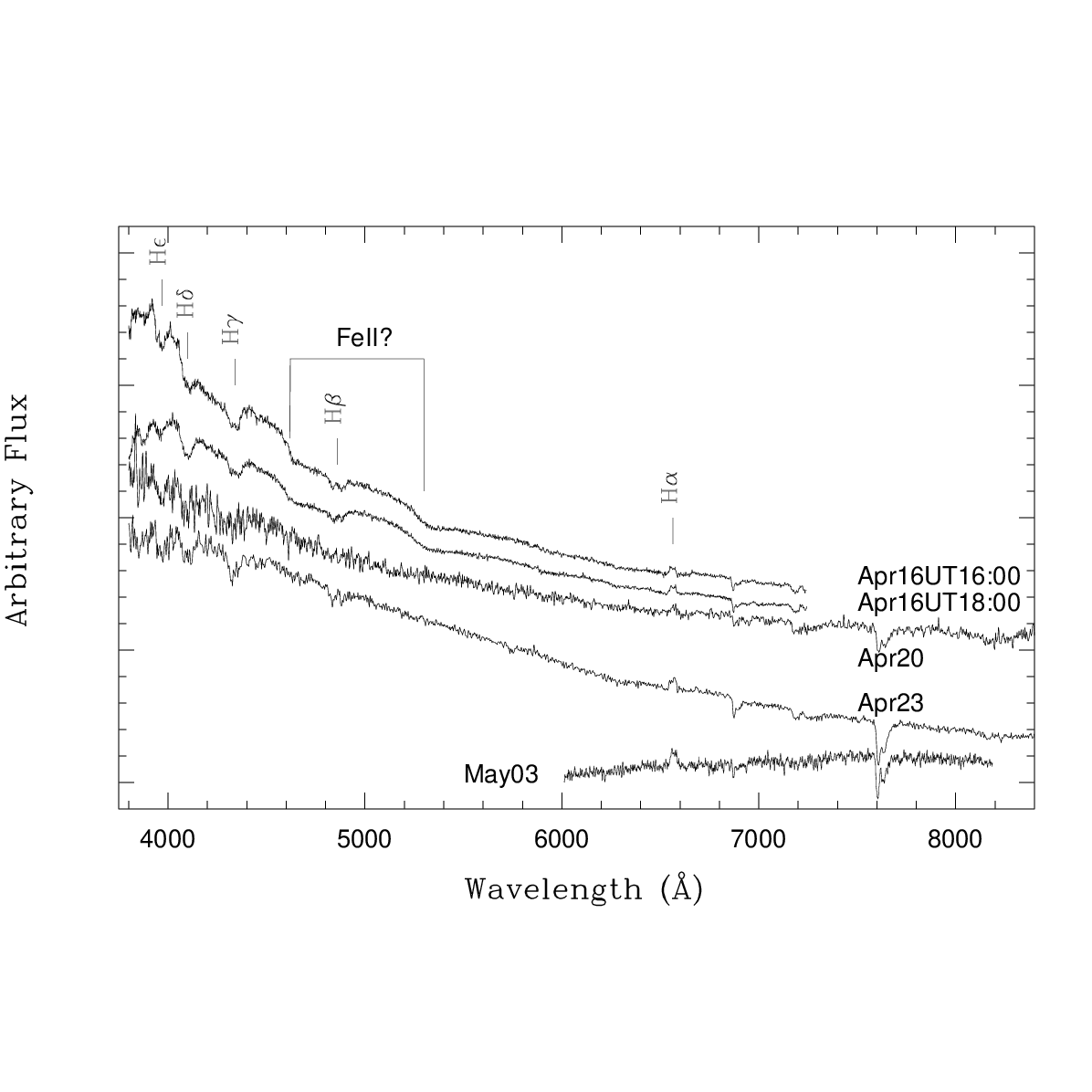}
\caption{Temporal spectral evolution of the transient GWAC\,180415A at the five different epochs from April 16 to May 03, 2018.
The spectra are shifted vertically by an arbitrary amount for visibility. 
The Balmer features, along with the possible \ion{Fe}{2}$\lambda$4600 and 5200 features, are marked on the 
top spectrum.
}
\end{figure}

\begin{figure}[ht!]
\plotone{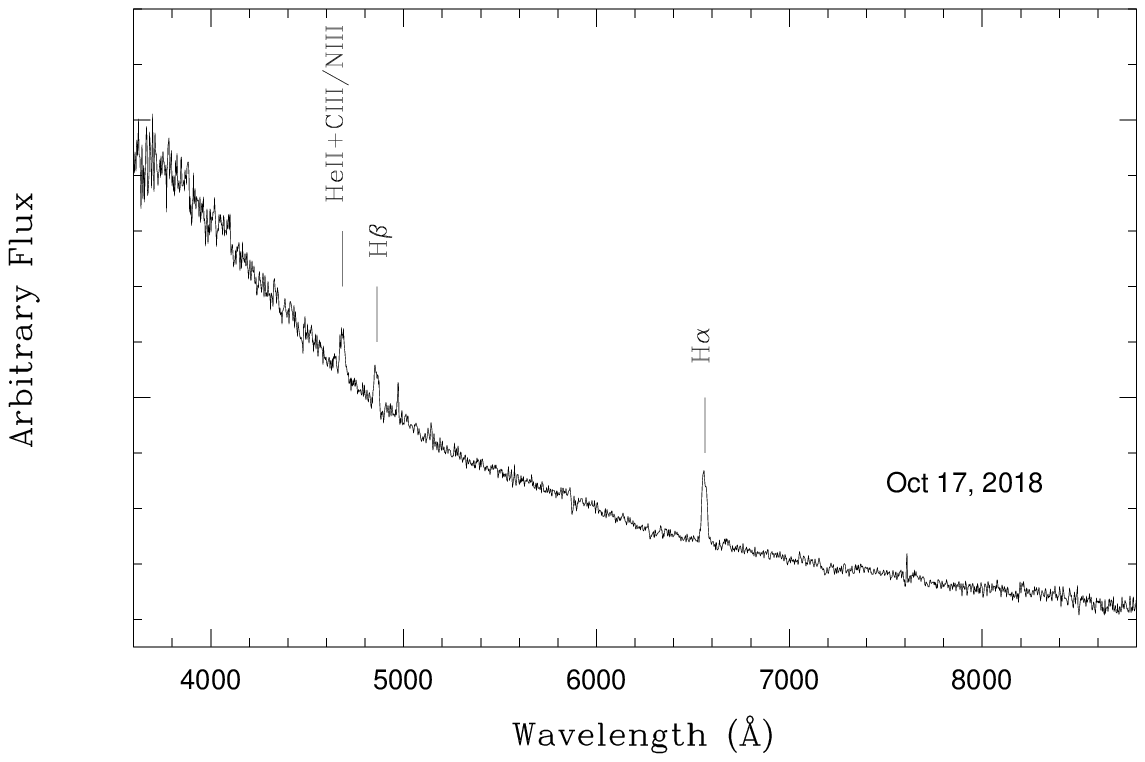}
\caption{The spectrum of GWAC\,181017A taken in October 17, 2018. 
The Balmer features, along with the \ion{He}{2}$\lambda4686$+\ion{C}{3}/\ion{N}{3}, are marked on the spectrum.
}
\end{figure}

\begin{figure}[ht!]
\plotone{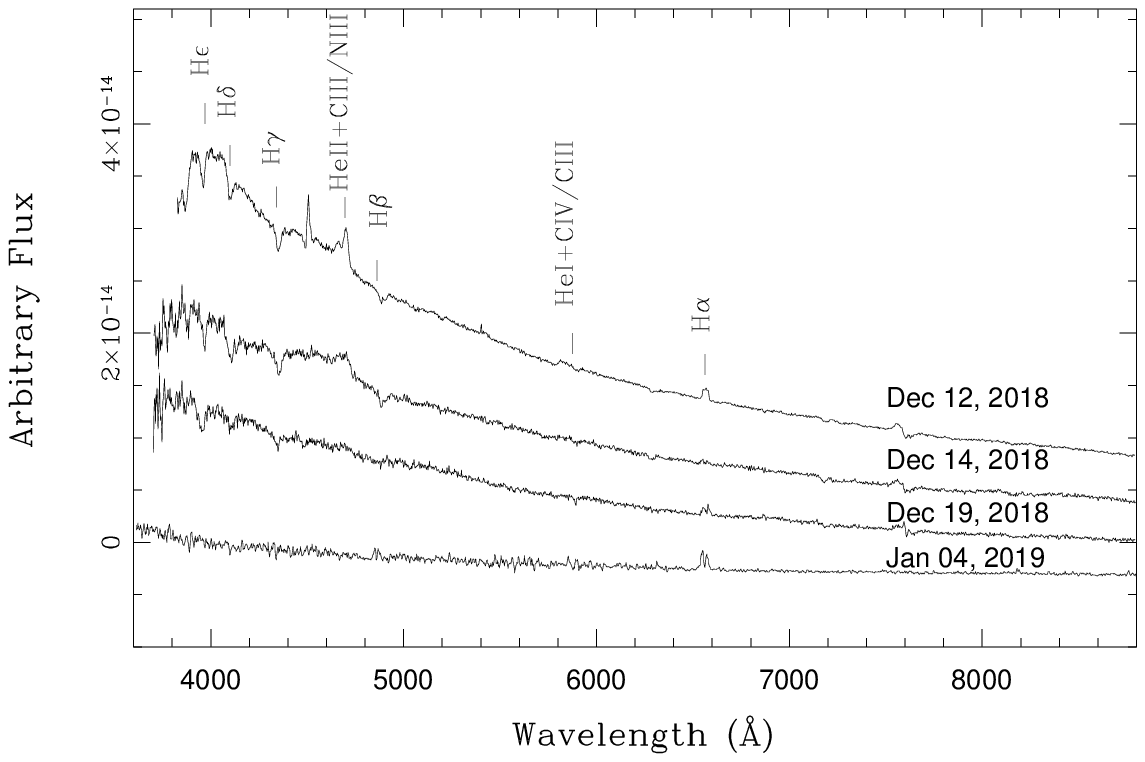}
\caption{The same as in Figure 4 but for GWAC\,181211A. In addition to the Balmer lines, the 
\ion{He}{2}$\lambda4686$+\ion{C}{3}/\ion{N}{3} and \ion{He}{1}$\lambda5016$+\ion{C}{4}/\ion{C}{3} features are marked 
on the top spectrum.
}
\end{figure}

\begin{figure}[ht!]
\plotone{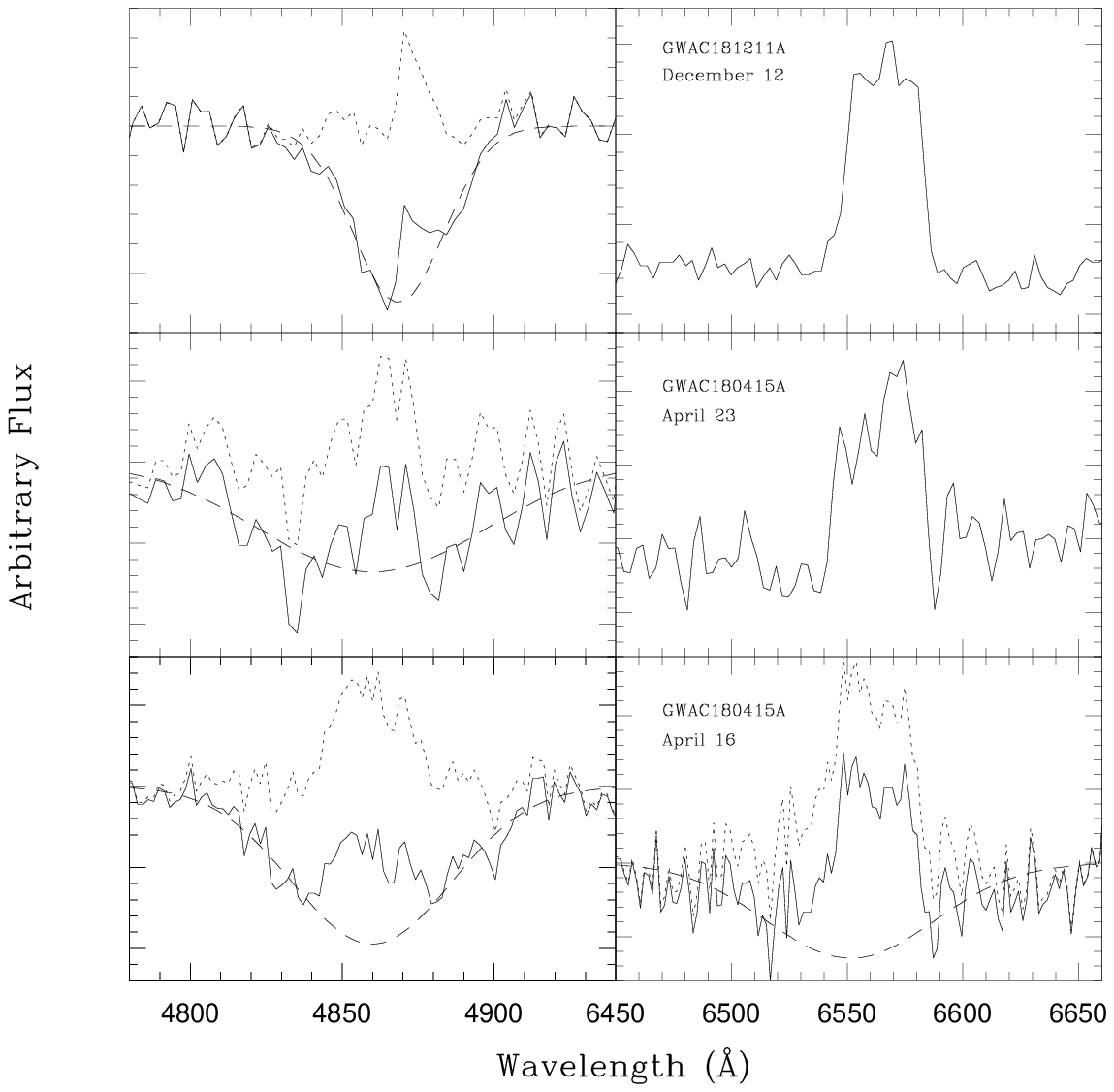}
\caption{The absorption corrected H$\beta$ (left panels) and H$\alpha$ (right panels) emission line profiles. In each panel, the 
observed profile and the modeled Gaussian absorption are plotted by the solid and long dashed lines, respectively. 
The short dashed line denotes the residual profile after removing the modeled absorption. 
}
\end{figure}





\subsection{Period Analysis}

All three superoutbursts exhibit an ordinary superhump in their light curves, which allows us to perform a period analysis to 
infer their basic physical parameters. The period analysis is carried out by using the PDM (Stellingwerf 1978)
task in the IRAF package. 

The left-upper panel in Figure 8 presents the early ($<0.25$ day) AAVSO light curve of 
GWAC\,180415A, revealing a transition from a raising stage to a declining stage. 
The upper-right panel shows the derived superhump period, $P = 0.0604$ day.
To obtain a more reliable period, we remove the underlying transition by
a cubic spline function which is overplotted
with the black solid line in the upper-left panel of Figure 8.
After removing the underlying transition (the left-lower panel), the superhump
period is determined to be $P=0.0563(1)$ day (the right-lower panel). 

The corresponding phase-averaged light curve is shown in the upper panel in 
Figure 9. One can see from the plot that there is a clear double-wave modulations, typical of the early superhump
usually detected in the early raising stage in a WZ Sge-type object (see Kato 2015 for a review). 
The existence of an early superhump whose period is extremely 
close to the orbital period has been, in fact, adopted as a modern definition of WZ Sge-type objects.   
The early superhump is well understood by the 2:1 resonance that is unique for the binary system with a mass ratio $q<0.1$ 
(e.g., Osaki \& Myer 2003). The phase-averaged light curve of the ordinary superhump, built from the latter light curve, 
is plotted in the lower panel in Figure 9. The superhump period of 0.0605 day is highly consistent with the
value of 0.057-0.060 days measured by Pavlenko et al. (2019).

Figures 10 and 11 present the phase-averaged light curves of the ordinary superhumps of GWAC\,181017A and GWAC\,181121A, respectively. 
For GWAC\,181121A, no early superhump period could be measured from the early-time light
curve provided by the GWAC's cameras, although the raising stage has been covered.

The resulted superhump period $P_{\mathrm{SH}}$ derived from our period analysis are given in Column (2) in Table 5. 
Without a measured early superhump period, we estimate the orbital period $P_{\mathrm{orb}}$ from the 
widely used relationship $\varepsilon=-3.3\times10^{-2}+0.84P_{\mathrm{SH}}/\mathrm{day}$ (Stolz \& Schoembs 1984), where $\varepsilon$ is the superhump excess defined as
$\varepsilon=(P_{\mathrm{SH}}-P_{\mathrm{orb}})/P_{\mathrm{orb}}$. 
For each superoutburst, the inferred $P_{\mathrm{orb}}$ and $\varepsilon$ are
listed in the first row of Column (3) and (4) in Table 5, respectively. For GWAC\,180415A, the values in brackets are inferred from 
the double-wave modulation identified in its early-time light curve.  
Because a few WZ-sge type objects with short $P_{\mathrm{orb}}$ and small $\varepsilon$ are found to be outliers of
the Stolz-Schoembs relationship, the second row presents the 
corresponding values obtained from a more modern relationship 
$P_{\mathrm{orb}}=0.91652(52)P_{\mathrm{SH}}+5.39(52)$ given by Gaensicke et al. (2009), 
where both $P_{\mathrm{orb}}$ and $P_{\mathrm{SH}}$ are in a unit of minute.

One can see from the table that both relationships return consistent results.
Both GWAC\,180415A and 
GWAC\,181211A have a small $\varepsilon\sim0.02$, while a large $\varepsilon\sim0.04$ can be found for GWAC\,181017A. 
Column (4) lists the mass ratio $q$ that is determined from the empirical relationship $\varepsilon=0.16q+0.25q^2$ 
(Patterson et al. 1998, 2005; Kato et al. 2009; Patterson 2011). 
Given their similar values of $\varepsilon$, both GWAC\,180415A and GWAC\,181211A are associated with 
a $q\approx0.1$, typical of WZ Sge-type DNe.  A larger $q\approx0.2$ is, however, revealed in 
GWAC\,181017A.

%
%

\begin{figure}[ht!]
\plotone{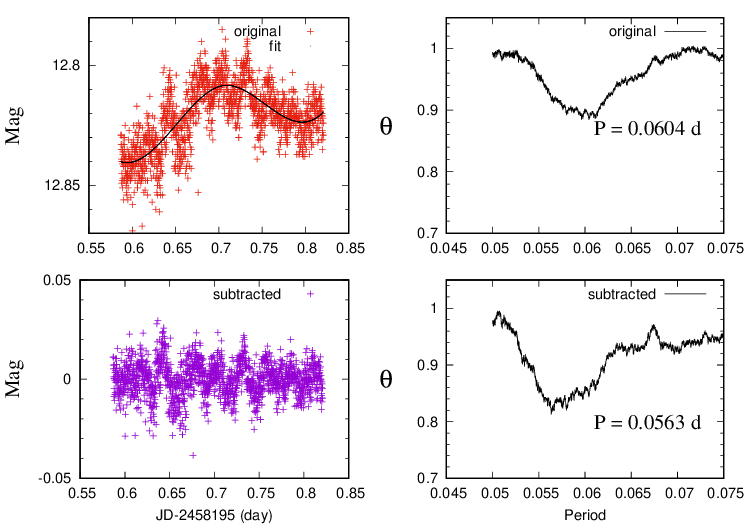}
\caption{A comparison of the period analysis on the early light curve of GWAC\,180415A. The left-upper panel shows the original 
light curve, and the right-upper panel the corresponding period analysis. The left-lower panel presents the 
light curve after a fitted sinusoidal function, which is overplotted on the left-upper panel by a solid line, is removed. 
The corresponding period analysis is illustrated in the right-lower panel. }
\end{figure}

\begin{figure}[ht!]
\plotone{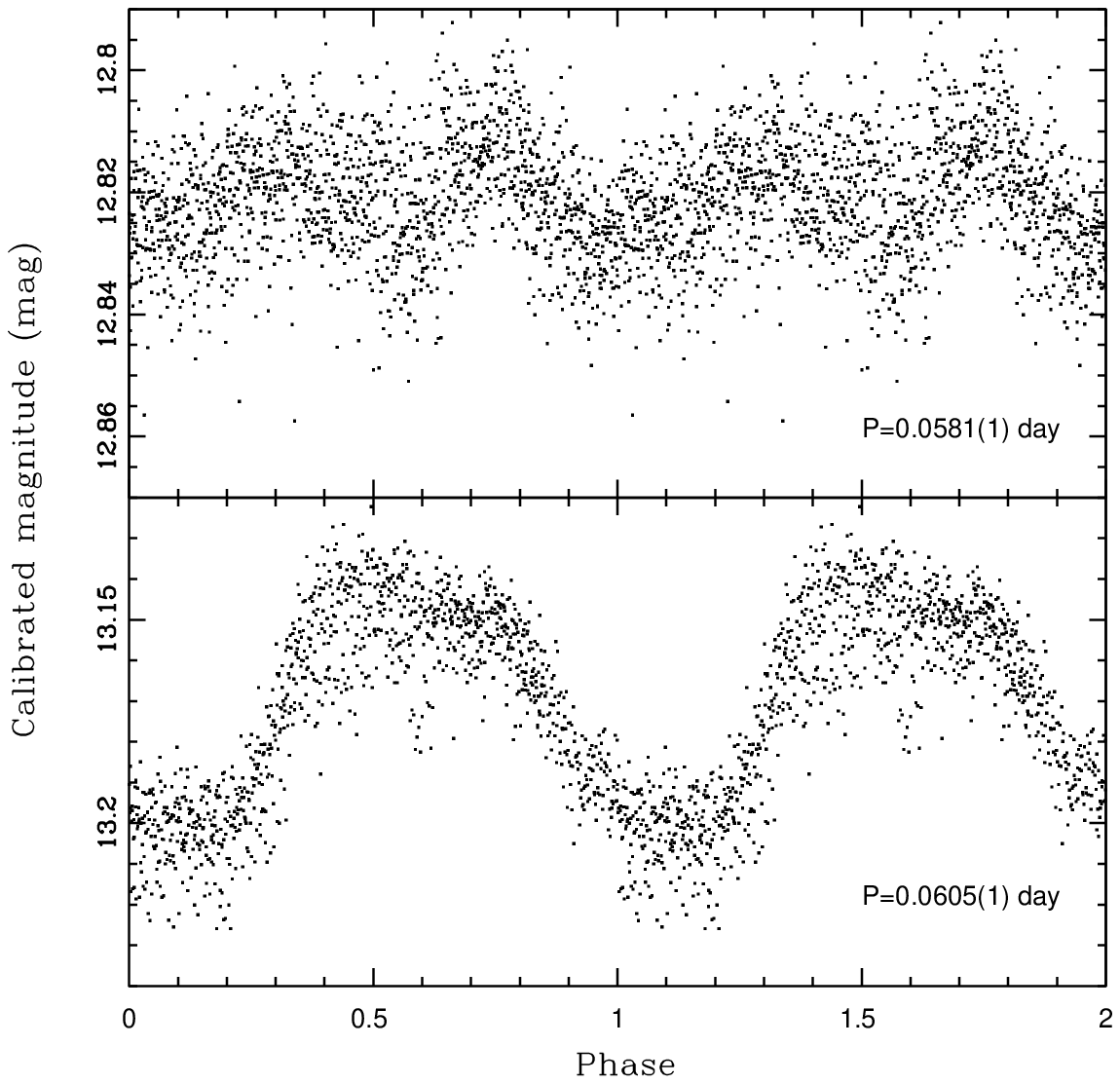}
\caption{\it Upper panel:\rm\ the averaged phase plot of the early light curve of GWAC\,180415A. \it Lower panel:\rm\ 
the same as the upper panel but for the late ordinary superhump. }
\end{figure}

\begin{figure}[ht!]
\plotone{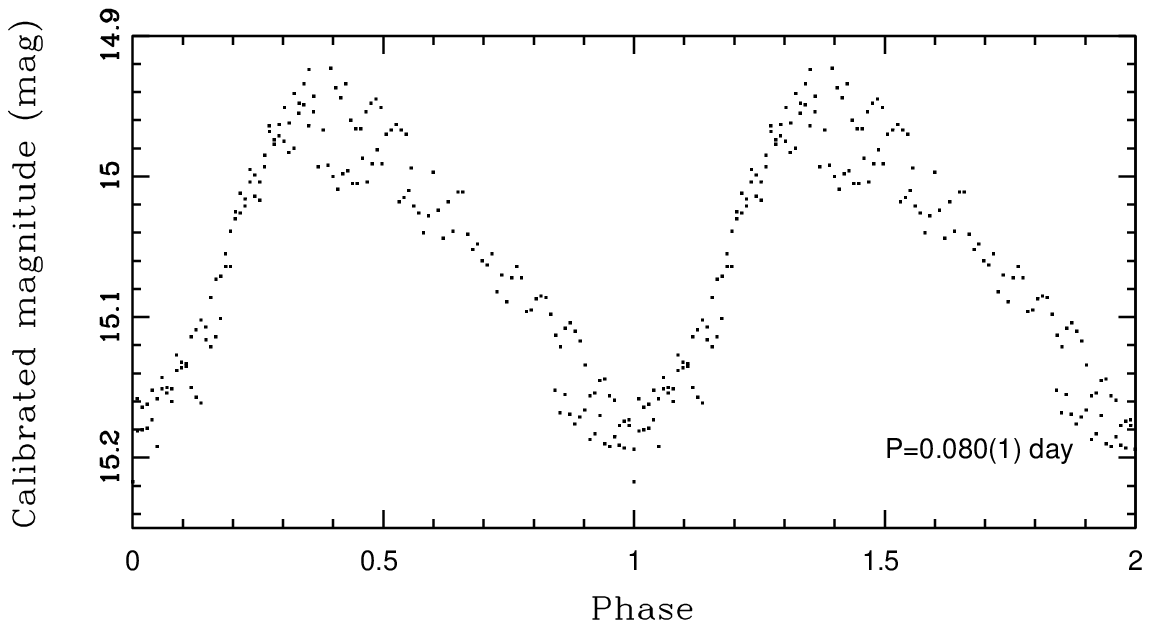}
\caption{The average phase plot of the superhump of GWAC\,181017A. }
\end{figure}

\begin{figure}[ht!]
\plotone{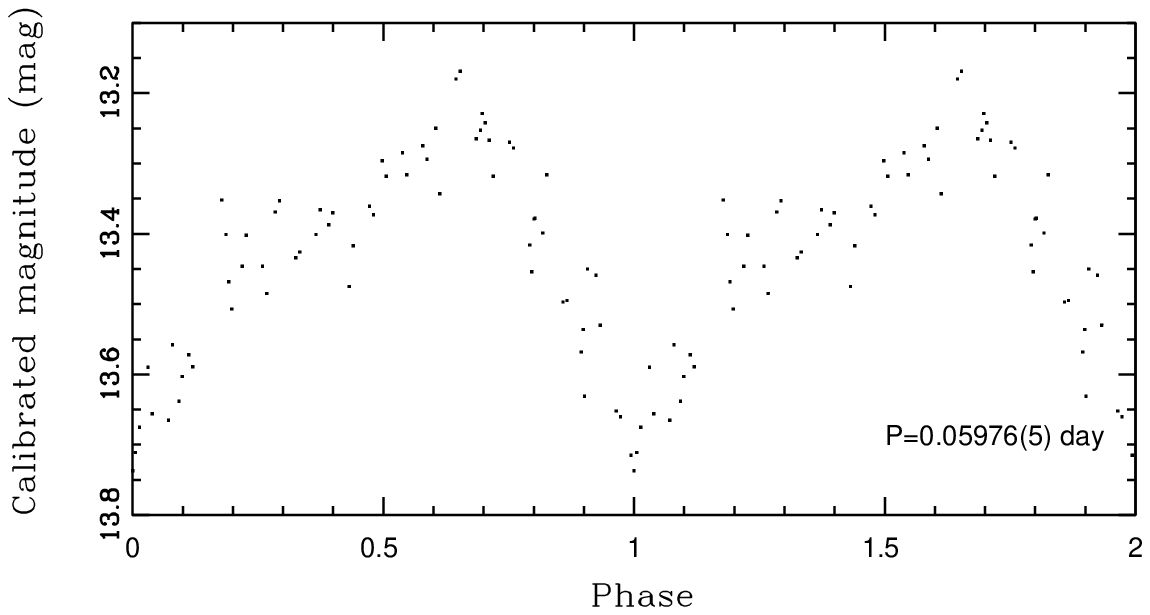}
\caption{The same as Figure 10 but for GWAC\,181211A. }
\end{figure}

\begin{table}[h!]
\renewcommand{\thetable}{\arabic{table}}
\centering
\caption{Period analysis of the three superoutbursts}
\label{tab:decimal}
\tiny
\begin{tabular}{ccccc}
\tablewidth{0pt}
\hline
\hline
Transients & $P_{\mathrm{SH}}$ & $P_{\mathrm{orb}}$ & $\varepsilon$ & $q$\\
           &   day   &   day  &  &   \\
 (1)  & (2) & (3) & (4) & (5) \\
\hline
GWAC\,180415A & 0.0605 & 0.0594(0.0563) &  0.0178(0.0746)  &  0.0967(0.313) \\
                           &             &  0.0592              &  0.0220                &  0.1163\\
GWAC\,181017A & 0.0800 & 0.0770              &  0.0389                &  0.1879\\
                            &             & 0.0770              &  0.0390                & 0.1883\\
GWAC\,181211A & 0.0598 & 0.0587              &  0.0181                &  0.0981\\
                            &             & 0.0585              & 0.0222                 &  0.1173\\
                                          
\hline
\hline
\end{tabular}
\end{table}

\section{Discussion}

We report our photometric and spectroscopic monitors of three DN superoutbursts independently detected by the GWAC system in 2018. 
Based on a combination of our follow-up observations and the data extracted from the AAVSO, 
our period analysis of the associated superhumps enables us to identify GWAC\,180415A and GWAC\,181211A as two new WZ Sge-type objects, 
both with small values of $\varepsilon<0.02$ and $q<0.1$.
Relatively large values of $\varepsilon\sim0.04$ and $q\sim0.2$ are found to be associated
with GWAC\,181017A. After correction of the absorption features, a relatively flat Balmer emission-line 
decrement H$\alpha$/H$\beta<2$ is 
revealed in GWAC\,180415A and GWAC\,181017A, and a large value of H$\alpha$/H$\beta=5.6$ in GWAC\,181211A.  

Although the normal outburst of a DN can be well understood by the disk instability (e.g., Warner 1995), a different origin is
necessary for the superoutburst. Patterson et al. (1981) and Osaki (1985) proposed that the superoutburst results from 
enhanced mass transfer;  however, this is inconsistent with the early superhump detected in the raising stage (Ishioka et al. 2002).
Osaki \& Kato (2013) instead suggests that superoutburst are caused by a thermal-tidal instability with constant
mass transfer.
A model with either an extremely low-$\alpha$ ($<0.003$) or a magnetically truncated disk is 
required to reproduce the long burst interval observed in the WZ Sge-type DNe,

We argue that the identification of GWAC\,180415A as a WZ Sge-type DN is additionally supported by 
several observed rebrightenings with a total duration of about two weeks,
although the exact mechanism of the rebrightening is still under debate. Proposed 
mechanisms include an enhanced $\alpha$ value due to magnetohydrodynamics (e.g., Osaki et al. 2001), 
accretion of the matter beyond the 3:1 resonance
(e.g., Kato et al. 1998), and repeat reflections of transverse waves in the disk (e.g., Meyer \& Meyer-Hofmeister 2015).

We estimate the mass of the secondary star in the three DNe from the inferred corresponding mass ratio $q$. 
The mean mass of WD ($M_{\mathrm{WD}}$) is $0.81\pm0.04M_\odot$ in CVs (Savoury et al. 2011), and $0.621M_\odot$ in the field SDSS magnitude-limited WD sample (Tremblay et al. 2016). 
With a range of $0.6M_\odot\leq M_{\mathrm{WD}}\leq0.8M_\odot$, the mass of the secondary is inferred 
to be $0.058M_\odot\leq M_2\leq0.094M_\odot$, $0.113M_\odot\leq M_2\leq0.153M_\odot$, and $0.073M_\odot\leq M_2\leq0.095M_\odot$ for GWAC\,180415A, GWAC\,181017A and GWAC\,181211A, respectively. 
Given the inferred masses of the secondary, we could not entirely exclude an existence of a BD in 
GWAC\,180415A and GWAC\,181211A.


The possible existence of a BD secondary in GWAC\,180415A is further supported by the detection of the quiescent 
counterpart in the IR by WISE (Cutri et al. 2013, and references therein),
although a contamination by a foreground or background source cannot be entirely excluded because of the low 
spatial resolution of WISE. 
Among the four IR bands, the object is detected only in the first two short bands:
$w1(3.4\micron)=17.17\pm0.13$ mag and $w2(4.6\micron)=16.22\pm0.20$ mag, implying an extremely red IR color 
of $w1-w2=0.95$ mag. There are no measurements in the $J$-, $H$-, and $K$-bands.

We first model the ultraviolet (UV) through IR spectral energy distribution (SED) of the quiescent counterpart of 
GWAC\,180415A  by a linear combination of two blackbodies through $\chi^2$ minimization,
deriving a WD temperature of $T_1 = 11, 400 \pm 500$K and a companion temperature of $T_2 = 950\pm40$K;
see Figure 12. The resulted reduced $\chi^2$ is 2.16 for 9 data points and 4 free parameters.  
We additionally model the SED by the theoretical grid of spectra of a pure hydrogen atmosphere given 
by Tremblay \& Bergeron (2009).  The surface gravity is fixed to be $\log g=8.0$ (cgs unit) 
in our modeling, which is the typical value of $\log g$ of the modeled atmospheric parameters of nearby WDs (Giammichele et al. 2012).  We at first fit the SED and calculate the corresponding $\chi^2$ statistics 
for a series of theoretical spectra with fixed effective temperature $T_{\mathrm{eff}}$. 
The matching is based on only the UV and optical bands. 
The best match with 
the lowest $\chi^2$ returns a consistent WD temperature $T_{\mathrm{eff}}=11,000$K.     
\rm
With the determined temperature and 
the assumed surface gravity, the WD mass is inferred to be $0.603M_\odot$ from the DA model grid calculated by 
Tremblay et al. (2011), suggesting a mass of the secondary of $0.058M_\odot$. 

\begin{figure}[ht!]
\plotone{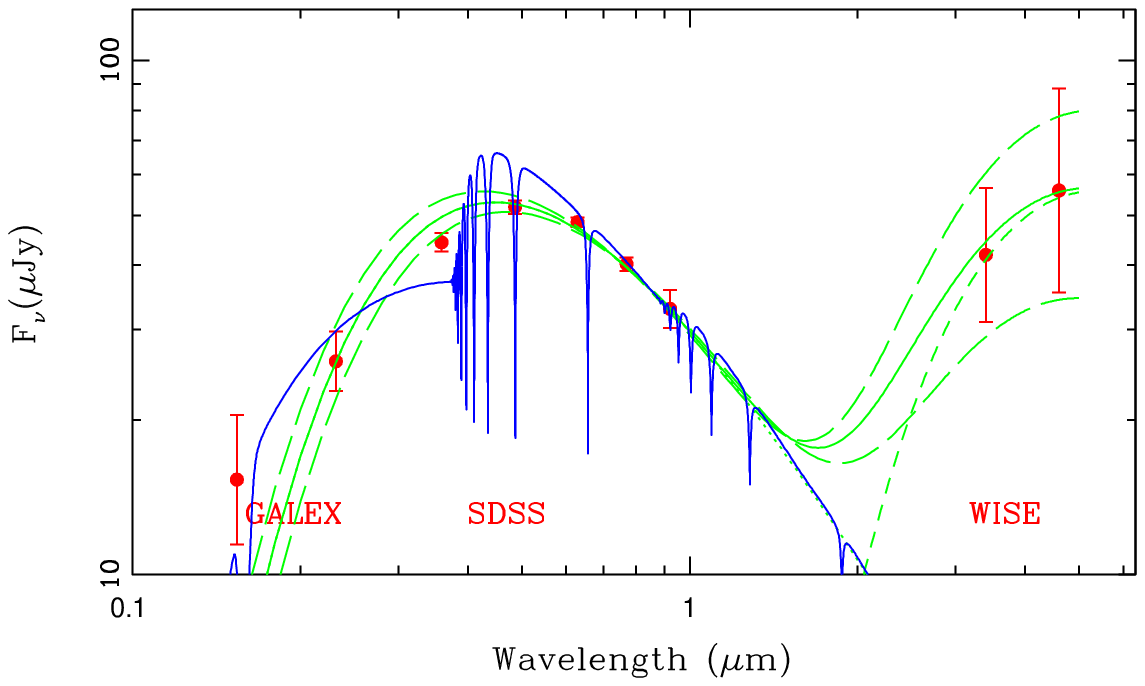}
\caption{The multi-wavelength SED (the red points) of SDSS\,J120857.3+191656.5, the quiescent counterpart of GWAC\,180415A, 
from FUV to infrared. The data are extracted from Galex (Martin et al. 2005), SDSS (Abazajian et al. 2009) and WISE (Cutri et al. 2013). 
The green solid and dashed lines show the best fit by
a linear combination of two blackbodies and the corresponding 1$\sigma$ confidence level. The blue line is the 
best fit theoretical spectrum with a temperature of $T =11,000$K that is extracted from the pure hydrogen grid built by 
Tremblay \& Bergeron (2009). 
}
\end{figure}

For the quiescent counterpart of GWAC\,181211A, we model the corresponding SDSS DR14 spectrum by the same theoretical spectral grid by the same method.
The surface gravity is again fixed to be $\log g=8.0$ since the modeling is found to be insensitive to the value of $\log g$
because of the poor S/N, which means there are only two free parameters.
 By ignoring the weak emission at the line centers of H$\alpha$ and H$\beta$ in 
the fitting, the best fit shows that the observed spectrum can be well reproduced by a model spectrum with $T=10,000$K, shown in Figure 13. The WD mass is therefore inferred to be 0.6$M_\odot$ through 
the same method, suggesting a mass of the secondary of $0.059M_\odot$.

\begin{figure}[ht!]
\plotone{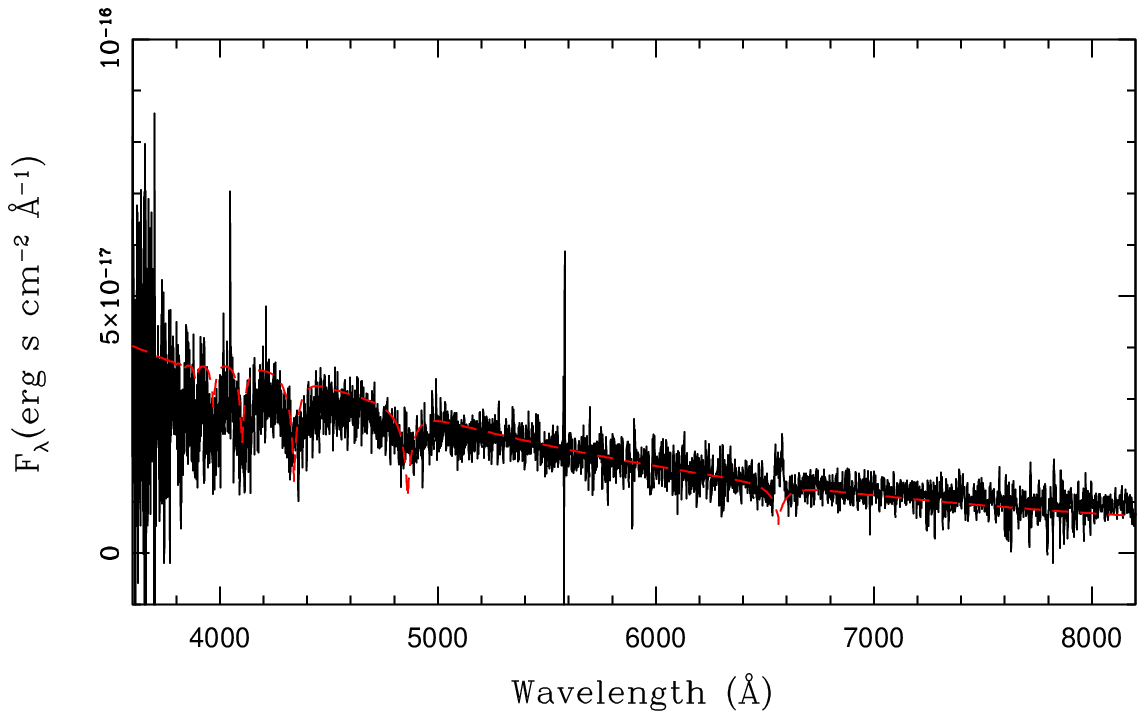}
\caption{An illustration of the modeling of the SDSS DR14 spectrum of SDSS\,J014823.29-004807.4, the quiescent counterpart of GWAC\,181211A.
The observed spectrum and the best fit DA model with a temperature of $T=10,000$K are shown by the black solid and red dashed curves, respectively.}
\end{figure}

Figure 14 shows the $P_{\mathrm{orb}}-\varepsilon (q)$ diagram that is a powerful tool for describing the evolution of 
a CV (e.g., Howell et al. 2001). Comparison of the observations and the theoretical evolution track 
with a $M_{\mathrm{WD}}=0.75M_\odot$ given by Knigge et al. (2011) confirms that both GWAC\,180415A and GWAC\,181211A 
are WZ Sge-type objects close to the period minimum because the locations of the two superoutbursts 
coincide with those of the 
previously confirmed WZ Sge-type objects. The comparison suggests that GWAC\,181017A is a prebouncer system.
We argue that this result is further supported by the detection of the quiescent counterpart by WISE. 
The quiescent counterpart of GWAC\,181017A  has $w1(3.4\micron)=17.19\pm0.13$ mag and $w2(4.6\micron)=17.29\pm0.54$ mag, giving an IR color of  $w1-w2=-0.1$ mag. This color differs from that of GWAC\,180415A significantly, implying a more massive secondary in GWAC\,181017A than in GWAC\,180415A. 
By extracting it from the Pan-STARRS1 surveys catalog (Chambers et al. 2016),
the more massive secondary can be additionally inferred from its red color, $r-i=0.31$ mag. 

\begin{figure}[ht!]
\plotone{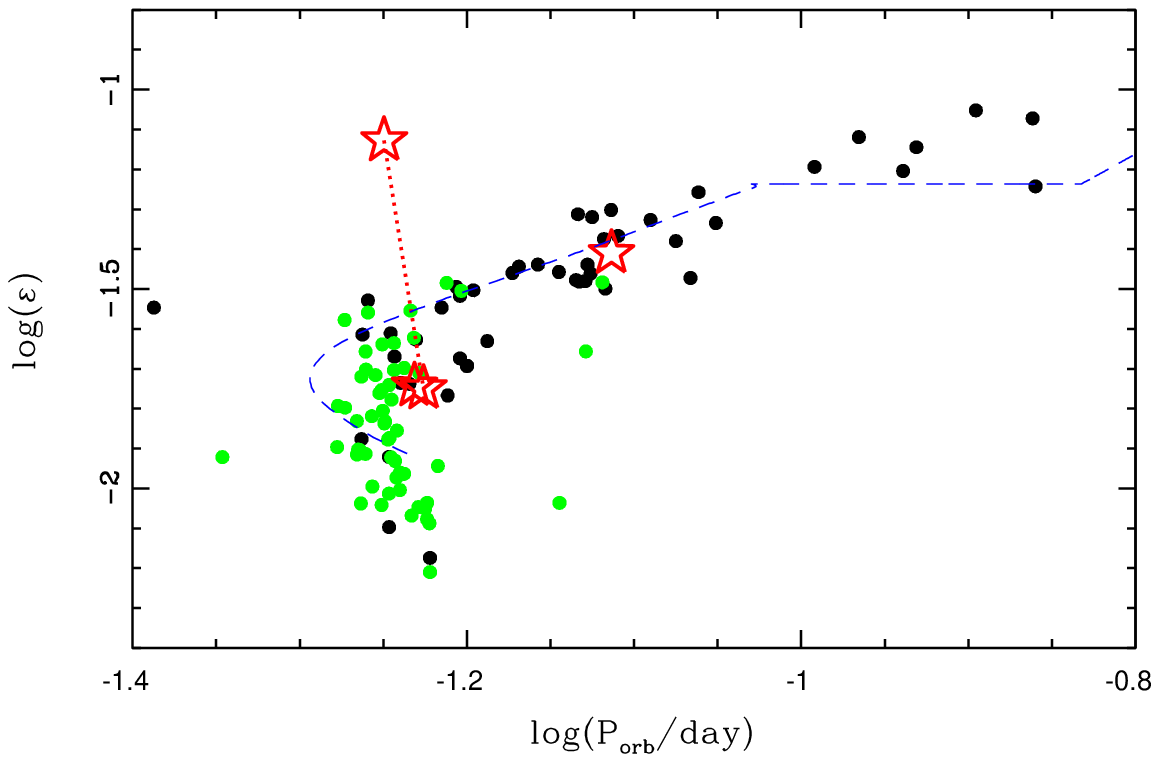}
\caption{The $\varepsilon-P_{\mathrm{orb}}$ diagram. The red stars mark the positions of the 
three superoutbursts studied in this paper. The two stars connected with a dashed line denotes the 
two measurements of GWAC\,180415A. One is based on the periods of both early light curve and late ordinary superhump, and the
another on the ordinary superhump period only. The black dots denote the pre-bouncer systems given in Patterson (2011), and 
the green ones the confirmed WZ Sge-type DNe taken from Kato (2015). The dashed blue line is the theoretically predicted 
evolutionary track of a CV with a WD mass of $0.75M_\odot$ provided in Kingge et al. (2011). }
\end{figure}


\section{Conclusion}

We report photometric and spectroscopic follow-up observations of the superoutbursts of three DNe
(GWAC\,180415A, GWAC\,181017A and GWAC\,181211A) that were
identified independently by ASAS-SN and the Ground Wide-angle Cameras system.
The mass ratios estimated from the period analysis of the ordinary superhumps are 
0.0967-0.1163, 0.1879-0.1883 and 0.0981-0.1173 for 
GWAC\,180415A, GWAC\,181017A and GWAC\,181211A (respectively). 
GWAC\,180415A shows not only long duration ($\sim2$ weeks) multiple rebrightenings with an 
amplitude of about 3-4 magnitudes, but also an early superhump associated with a double-wave modulation, which 
indicates a firm identification of the object as a  WZ sge-type DNe. GWAC\,181211A is a WZ sge-type DNe candidate
taking into account of its small mass ratio.
The measured Balmer decrements suggest the Balmer line emission is produced from an optical 
thick region in GWAC\,180415A and GWAC\,181017A, and from an optical thin region in GWAC\,181211A. 

\rm
\acknowledgments

The authors thank the anonymous referee for his/her careful review and helpful suggestions that improved the
manuscript.
The study is supported by the National Basic Research
Program of China (grant 2014CB845800), the NSFC under grants 11533003, and the Strategic
Pioneer Program on Space Science, Chinese Academy of
Sciences, grant Nos. XDA15052600 \& XDA15016500.  
JW is supported by the National Natural Science Foundation of China under grant 11773036, 
by Natural Science Foundation of Guangxi (2018GXNSFGA281007), and by Bagui Young Scholars Program, 
JM is supported by the NSFC grants 11673062, the Hundred Talent Program of Chinese Academy of Sciences, 
the Major Program of Chinese Academy of Sciences (KJZD-EW-M06), and the Oversea Talent Program of Yunnan Province.
Special thanks go to the staff at Xinglong Observatory
as a part of National Astronomical Observatories, China
Academy of Sciences for their instrumental and observational help. 
This study is supported by the Open Project Program of
the Key Laboratory of Optical Astronomy, NAOC, CAS. 
The GWAC system is partially funded by National Astronomical Observatories, CAS, and the Guangxi Key Laboratory for Relativistic 
Astrophysics. 
This study
used the SDSS archive data that was created and distributed by
the Alfred P. Sloan Foundation, the Participating Institutions,
the National Science Foundation, and the U.S. Department of
Energy Office of Science.
This study uses the data collected by Wide-field Infrared
Survey Explorer (WISE), which is a joint project of the University of California, Los Angeles, and the Jet 
Propulsion Laboratory/California Institute of Technology, funded by the National Aeronautics and Space Administration.
This study uses the data obtained by GALEX, which is a NASA Small Explorer, and uses  
the NASA/IPAC Extragalactic Database (NED), which is operated by the Jet Propulsion 
Laboratory, California Institute of Technology, under contract with NASA.
This work has made use of data from the European Space Agency (ESA) mission
{\it Gaia} (\url{https://www.cosmos.esa.int/gaia}), processed by the {\it Gaia}
Data Processing and Analysis Consortium (DPAC,
\url{https://www.cosmos.esa.int/web/gaia/dpac/consortium}). Funding for the DPAC
has been provided by national institutions, in particular the institutions
participating in the {\it Gaia} Multilateral Agreement.

\facilities{Ground Wide-Angle Cameras (GWAC), GWAC-F60A/B telescope, Xinglong Observatory 2.16m telescope,
Lijiang 2.4m telescope (LJT), Shane 3m telescope, 0.76 m Katzman Automatic Imaging Telescope (KAIT)}

\software{IRAF (Tody 1986, 1993), Python, IDL}

\appendix
\section{Designation of GWAC system}

A refractive telescope with an effective aperture size of 18 cm is adopted for each camera.
A field corrector composed of 8 lenses is designed for each camera to correct the comet aberration at the 
edge of the FoV.
The $f$-ratio is $f/$1.2 for each camera, resulting in a FoV of $150\mathrm{deg^2}$ and a spatial resolution of 
11.7\arcsec$\cdot\mathrm{pixel^{-1}}$ when 
each of the cameras is equipped with a 4096$\times$4096 E2V back-illuminated CCD chip
operating in the 0.5 to 0.85 $\mu$m band. 
All the CCDs with a full well of about $10^5e^-$ are operated at a gain of about 2$e^{-}\ \mathrm{ADU^{-1}}$.
The readout noise is 13$e^{-}\ \mathrm{pixel^{-1}}$ for each chip when the readout time is set to 5 seconds.
The thermoelectric coolers can keep the CCDs at a temperature of -50\degr C, which is acceptable because 
at this temperature the noise of a broad-band and large-area  sky survey is mainly contributed by the sky background rather than the dark current. 
No filters are used in order to have a relatively deep detection capability.
In a 10 s exposure,
limiting magnitudes of $m_V=15$ and $m_V=16$ mag\footnote{The white-band magnitude can be well calibrated and transformed to the $R-$band in the Johnson system. 
In order to be consistent with the requirement of GWAC system presented in the white paper (Wei et al. 2016), 
the corresponding $V-$band magnitude is obtained by transfroming from the calibrated $R-$band magnitude by assuming a spectrum of a G-type star.}   
%
%
can be reached at a significance of 5$\sigma$ during 
full moon and new moon, respectively, and 
the cameras saturate for objects brighter than $m_\mathrm{V}=11$ mag.

The GWAC cameras adopt a German equatorial mount tracking in both axes,  a superior choice for an all-sky survey. 
Each mount carries four cameras (called a ``unit'' of the GWAC system) pointing to different
sky areas, so the total FoV of each unit is $\sim600\ \mathrm{deg^2}$.
The maximum slew velocity of the mounts is $1.5\arcdeg\ \mathrm{s^{-1}}$ for both axes.
The tracking accuracy over 5 minutes is better than 2\arcsec\ at the zenith.
A closed loop based on the astrometry of the detected stars is used to guarantee high tracking accuracy during a
long-duration observation.

\section{Survey strategy}

As a start point, we divide the whole sky into several grids having equivalent areas that correspond to the FoV of one GWAC unit. In the survey,  
each GWAC unit is assigned to a given grid. The monitor duration is from
half an hour to as long as six hours for an individual grid.
When making the observing schedule,  the grids with a Galactic latitude of $b<20$\degr\ are avoided, 
since the extremely high star density in the GWAC images greatly reduces the detection efficiency of any transient.
During the survey,  a minimum angular distance from the center of each grid to the Moon is 
necessary for avoiding both strong sky background and bright stray light recorded in the GWAC images.
The minimum distance depends on the lunar phase: 30\degr when the phase is less than 0.5, 
40\degr when the phase is 0.5-0.75, and 40\degr\ when the phase is $>$0.75.

Based on the catalog cross-match method, a dedicated real-time transient detection pipeline has been developed for the GWAC system.
Briefly, all the point sources detected in each GWAC image are compared to a reference catalog to find
transient candidates. The reference catalog is produced by combining the constantly updated local GWAC catalog and the available catalogs 
provided by other whole sky surveys such as the USNO B1.0.
The local GWAC catalog is built from stacked, high-quality reference images taken by the 
GWAC system in advance. 
This strategy based on a combination of catalogs has advantages. On the one hand, the public astronomical catalogs are static, 
which would result in heavy pollution from false positives such as the known long-term variable stars (e.g., Mira variable stars). 
On the other hand, although the false positive issue can be somewhat addressed by the local GWAC catalog,
the completeness of the catalog is reduced in some cases. 
For example, some faint objects around very bright stars are difficult to be found in the GWAC images with
low spatial resolution, and they are consequently excluded from the local GWAC catalog.

We note that the combination of the local GWAC catalog and the archive catalogs must be finely tuned, since the 
density of stars in each field is different. 
A list of proper and auto-adapted parameters should be set to ensure that the used reference catalog not only is 
complete at
the sensitivity limit of the GWAC cameras, but also does not contain too many redundant faint objects.
For the GWAC system, the reference images and the reference catalog are updated as long as new images with higher quality are obtained. 

In addition to real optical transients, the transient candidates detected by the GWAC detection pipeline include
asteroids, meteors, satellites, comets, airplanes, hot pixels, dust on the mirror or CCD, and image defects.
Algorithms are involved in our pipeline to filter out these false-positives and contamination.
Some of them, such as satellites, can be efficiently removed by analyzing a series of more than two consecutive images. 
A pipeline based on machine-learning algorithm has been developed to filter out the variable hot pixels and dust (Xu et al. in preparation). 
For slowly moving objects (e.g., minor planets), custom-developed software has been running in 
the local database to predict the positions and brightnesses for all the known minor planets. The prediction 
can be cross-matched with the observed sky position of a transient candidate in real time when a new alert is generated.  
The accuracy of this dedicated software are better than 0.5\arcmin, which has been checked with the predictions 
given by MPC\footnote{https://www.minorplanetcenter.net/} on-line. 



\begin{thebibliography}{}

\bibitem[Abazajian et al. (2009)]{aba09} Abazajian, K. N., Adelman-McCarthy, J. K., Agueros, M. A., et al. 2009, \apjs, 182, 543
\bibitem[Aviles et al. (2010)]{avi10} Aviles, A., Zharikov, S., \& Tovmassian, G. 2010, \apj, 711, 389
\bibitem[Baba et al. (2002)]{bab02} Baba, H., Sadakane, K., Norimoto, Y.,  et al. 2002, \pasj, 54, L7
\bibitem[Breedt et al. (2014)]{bre14} Breedt, E., Gansicke, B. T., Drake, A. J., et al. 2014, \mnras, 443, 3174
\bibitem[Burd et al. (2015)]{bur15} Burd, A., Cwiok, M., Czyrkowski, H., et al. 2015, \na, 10, 409
\bibitem[Chambers et al. (2016)]{cha16} Chambers, K. C., Magnier, E. A., Metcalfe, N., et al. 2016, arXiv:astro-ph/1612.05560
\bibitem[Ciardi et al. (1998)]{cia98} Ciardi, D. R., Howell, S. B., Hauschildt, P. H., et al. 1998, \apj, 504, 450
\bibitem[Cutri et al. (2013)]{cut13} Cutri, R. M., et al. 2013, yCat, 2328
\bibitem[Fan et al. (2015)]{fan15} Fan, Yu-Feng, Bai, Jin-Ming, Zhang, Ju-Jia, Wang, Chuan-Jun, Chang, Liang, Xin, Yu-Xin, \& Zhang, Rui-Long
2015, RAA, 15, 918
\bibitem[Fan et al. (2016)]{fan16} Fan, Z., Wang, H. J., Jiang, X. J., et al. \pasp, 128, 5005
\bibitem[Gaia Collaboration et al. (2018)]{gai18} Gaia Collaboration, et al. 2018, \aap, 616, 1
\bibitem[Ganeshalingam et al. (2010)]{gan10} Ganeshalingam, M., Li, W. D., Filippenko, A. V., et al. 2010, \apjs, 190, 418
\bibitem[Gansicke et al. (2009)]{gan19} Gansicke, B. T., Dillon, M., Southworth, J., et al. 2009, \mnras, 397, 2170
\bibitem[Goliasch \& Nelson (2015)]{gln15} Goliasch, J. \& Nelson, L. 2015, \apj, 809, 80
\bibitem[Graham et al. (2019)]{gra19} Graham, M. J., Kulkarni, S. R., Bellm, E. C., et al. 2019, \pasp, 131, 1001
\bibitem[Hiroi et al. (2009)]{hir09} Hiroi, K., Moritani, Y., Nogami, D., et al. 2009, \pasj, 61, 697
\bibitem[Howell \& Ciardi (2001)]{hoc01} Howell, S. B., \& Ciardi, D. R. 2001, \apjl, 550, 57 
\bibitem[Howell et al. (1997)]{how97} Howell, S. B., Rappaport, S., \& Politano, M. 1997, \mnras, 287, 929
\bibitem[Howell et al. (2001)]{how01} Howell, S. B., Nelson, L. A., \& Rappaport, S.  2001, \apj, 550, 897 
\bibitem[Imada et al. (2006)]{ima06} Imada, A., Kubota, K., Kato, T., et al. 2006, \pasj, 58, L23
\bibitem[Ishioka et al. (2002)]{ish02} Ishioka, R., Uemura, M., Matsumoto, K., et al. 2002, \aap, 381, L41 
\bibitem[Ivezic et al. (2008)]{ive08} Ivezic, Z., Axelrod, T., Becker, A. C., et al. 2008,  AIP Conference Proceedings, 1082, 359
\bibitem[Kato (2015)]{kat15} Kato, T. 2015, \pasj, 67, 108
\bibitem[Kato et al. (2009a)]{kat09a} Kato, T., Imada, A., Uemura, M., et al. 2009, \pasj, 61, 395
\bibitem[Kato et al. (2009b)]{kat09b} Kato, T., Pavlenko, E. P., Maehara, H., 2009, \pasj, 61, 601
\bibitem[Kato et al. (2012)]{kat12} Kato, T., Maehara, H., Miller, I., et al. 2012, \pasj, 64, 21
\bibitem[Kato et al. (1998)]{kat98} Kato, T., Nogami, D., Masuda, S., \& Baba, H. 1998, \pasp, 110, 1400
\bibitem[Kolb (1993)]{kol93} Kolb, U. 1993, \aap, 271, 149
\bibitem[Knigge et al. (2011)]{kni11} Knigge, C., Baraffe, I., \& Patterson, J. 2011, \apjs, 194, 28 
\bibitem[Kulkarni (2018)]{kul18} Kulkarni, S. R. 2018, ATel, 11266, 1
\bibitem[Lasota (2001)]{las01} Lasota, J. -P. 2001, \nar, 45, 449
\bibitem[Longstaff et al. (2019)]{lon19} Longstaff, E. S., Casewell, S. L., Wynn, G. A., Page, K. L., Williams, P. K. G., Braker, I., \& 
Maxted, P. F. L. 2019, \mnras, 484, 2566
\bibitem[Lin \& Papaloizou (1979)]{lip79} Lin, D. N. C., \& Papaloizou, J. 1979, \mnras, 186, 799 
\bibitem[Littlefair et al. (2000)]{lit00} Littlefair, S. P., Dhillon, V. S., Howell, S. B., \& Ciardi, D. R. 2000, \mnras, 313, 117
\bibitem[Littlefair et al. (2003)]{lit03} Littlefair, S. P., Dhillon, V. S., \& Martín, E. L. 2003, \mnras, 340, 264 
\bibitem[Littlefair et al. (2013)]{lit13} Littlefair, S. P., Savoury, C. D. J., Dhillon, V. S., et al. 2013, \mnras, 431, 2820
\bibitem[Martin et al. (2005)]{mar05} Martin, D. C., Fanson, J., Schiminovich, D., et al. 2005, \apjl, 619, 1
\bibitem[Massey et al. (1998)]{mas98} Massey, P., Strobel, K., Barnes, J. V., et al. 1988, \apj, 328, 315
\bibitem[McAllister et al. (2017)]{mva17} McAllister, M. J., Littlefair, S. P., Dhillon, V. S., et al. 2017, \mnras, 467, 1024
\bibitem[Mennickent et al. (2004)]{men04} Mennickent, R. E., Diaz, M. P., \& Tappert, C. 2004, \mnras, 347, 1180  
\bibitem[Meyer \& Meyer-Hofmeister (1981)]{mem81} Meyer, F., \& Meyer-Hofmeister, E. 1981, \aap, 104, L10
\bibitem[Meyer \& Meyer-Hofmeister (2015)]{mem15} Meyer, F., \& Meyer-Hofmeister, E. 2015, \pasj, 67, 52 
\bibitem[Negueruela et al. (2000)]{neg00} Negueruela, I., Reig, P, \& Clark, J. S. 2000, \aap, 345, L29
\bibitem[Neustroev et al. (2017)]{neu17} Neustroev, V. V., Marsh, T. R., Zharikov, S. V., et al. 2017, \mnras, 467, 597
\bibitem[Osaki (1985)]{osk85} Osaki, Y. 1985, \aap, 144, 369
\bibitem[Osaki (1989)]{osa89} Osaki, Y. 1989, \pasj, 41, 1005
\bibitem[Osaki (1996)]{osa96} Osaki, Y. 1996, \pasp, 108, 39
\bibitem[Osaki \& Kato (2013)]{osk13} Osaki, Y., \& Kato, T. 2013, \pasj, 65, 95 
\bibitem[Osaki \& Meyer (2003)]{osm03} Osaki, Y., \& Meyer, F. 2003, \aap, 401, 325
\bibitem[Osaki et al. (2001)]{osk01} Osaki, Y., Meyer, F., \& Meyer-Hofmeister, E. 2001, \aap, 370, 488 
\bibitem[Pala et al. (2018)]{pal18} Pala, A. F., Schmidtobreick, L., Tappert, C., et al. 2018, \mnras, 481, 2523
\bibitem[Patterson (2011)]{pat11} Patterson, J. 2011, AAS, 21810303 
\bibitem[Patterson et al. (2005)]{pat05} Patterson, J., Kemp, J., Harvey, D. A., et al. 2005, \pasp, 117, 1204
\bibitem[Patterson et al. (2002)]{pat02} Patterson, J., Masi, G., Richmond, M. W., et al. 2002, \pasp, 114, 721
\bibitem[Patterson et al. (1981)]{pat81} Patterson, J., McGraw, J. T., Coleman, L., \& Africano, J. L. 1981, \apj, 248, 1067 
\bibitem[Patterson et al. (1998)]{pat98} Patterson, J., Richman, H., Kemp, J., \& Mukai, K. 1998, \pasp, 110, 403
\bibitem[Ritter \& Kolb (2003)]{rik03} Ritter, H., \& Kolb, U. 2003, \aap, 404, 301
\bibitem[Savoury et al. (2011)]{sav11} Savoury, C. D. J., Littlefair, S. P., Dhillon, V. S., et al. 2011, \mnras, 415, 2025
\bibitem[Shappee et al. (2014)]{sha14} Shappee, B., Prieto, J., Stanek, K. Z., et al. 2014, AAS, 223, 23603
\bibitem[Sheets et al. (2007)]{she07} Sheets, H. A., Thorstensen, J. R., Peters, C. J., \& Kapusta, A. B. 2007, \pasp, 119, 494
\bibitem[Sarty \& Wu (2006)]{saw06} Sarty, G. E., \& Wu, K. 2006, \pasa, 23, 106
\bibitem[Stahl et al. (12019)]{sta19} Stahl, B., Zheng, W., de Jaeger, T., et al., 2019, arXiv:1909.11140
\bibitem[Stolz \& Schoembs (1984)]{stx84} Stolz, B., \& Schoembs, R. 1984, \aap, 132, 187 
\bibitem[Stellingwerf (1978)]{ste78} Stellingwerf, R. F. 1978, \apj, 221, 661 
\bibitem[Tody (1986)]{tod86} Tody, D. 1986, Proc. SPIE, 627, 733
\bibitem[Tody (1993)]{tod93} Tody, D. 1993, in ASP Conf. Ser. 52, adass II, ed. R. J.
\bibitem[Tonry et al. (2018)]{Patterson, Joseph; Richman, Hayley; Kemp, Jonathan; Mukai, Kojiton18} Tonry, J. L., Denneau, L., Heinze, A. N., et al. 2018, \pasp, 130, 4505
\bibitem[Tremblay \& Bergeron (2009)]{trb09} Tremblay, P. -E., \& Bergeron, P. 2009, \apj, 696, 1755
\bibitem[Tremblay et al. (2011)]{tre11} Tremblay, P. -E., Bergeron, P., \& Gianninas, A. 2011, \apj, 730, 128
\bibitem[Tremblay et al. (2016)]{tre16} Tremblay, P. -E., Cummings, J., Kalirai, J. S., Gansicke, B. T., Gentile-Fusillo, N., \& Raddi, R. 2016, \mnras, 461, 2100
\bibitem[Turpin et al. (2019)]{tur19} Turpin, D., Wu, C., Han, X. H., et al. 2019, RAA submitted, arXiv:astro-ph/1902.08476 
\bibitem[Vestrand et al. (2004)]{×} Vestrand, W. T., Borozdin, K. N., Brumby, S. P., et al. 2004, SPIE, 4845, 126
\bibitem[Vogt (1982)]{vog82} Vogt, N. 1982, \apj, 252, 653
\bibitem[Warner (1995)] {war95} Warner, B. 1995, Cataclysmic Variable Stars, Cambridge Astrophisical Ser. 28; Cambridge Univ. Press, Cambridge
\bibitem[Wei et al. (2016)]{wei16} Wei, J. Y., Cordier, B., et al. 2016, arXiv:astro-ph/1610.0689
\bibitem[Whitehurst (1988)]{whi88} Whitehurst, R. 1988, \mnras, 232, 35
\bibitem[Williams (1995)]{wil95} Williams, G. A., 1995, \aj, 109, 319
\bibitem[Wu et al. (2001)]{wu01} Wu, X. A., Li, A. Y., \& Gao, W. H. 2001, \apjl, 549, 81
\bibitem[Wu et al. (1995)]{wu95} Wu, K., Wickramasinghe, D. T., \& Warner, B. 1995, \pasa, 12, 60
\bibitem[Zhao et al. (2006)]{zha06} Zhao, Y. H., Li, Z. Y., Wu, X. A., \& Peng, Q. H. 2006, \aj, 131, 1667














\end{thebibliography}
\end{document}